\documentclass[journal,comsoc]{IEEEtran}

\IEEEoverridecommandlockouts
\usepackage{cite}
\usepackage{amsmath,amssymb,amsfonts}

\usepackage{graphicx}
\usepackage{textcomp}
\usepackage{xcolor}

\usepackage{algpseudocode}
\usepackage{algorithm}
\usepackage{algorithmicx}

\usepackage{balance}
\usepackage{enumitem}
\usepackage{color,soul}

\usepackage{mathtools}
\DeclarePairedDelimiter\ceil{\lceil}{\rceil}
\DeclarePairedDelimiter\floor{\lfloor}{\rfloor}
\DeclarePairedDelimiter{\nint}\lfloor\rceil

\usepackage{booktabs,makecell,tabularx}

\newcommand{\vect}{\operatorname{vect}}
\newcommand{\E}{\operatorname{E}} 
\newcommand{\tr}{\operatorname{tr}} 

\def\BibTeX{{\rm B\kern-.05em{\sc i\kern-.025em b}\kern-.08em
    T\kern-.1667em\lower.7ex\hbox{E}\kern-.125emX}}
    
\usepackage{fancyhdr}
\fancypagestyle{specialfooter}{
  \fancyhf{}
  
  \lhead{\small Accepted for publication in \textit{IEEE Transactions on Vehicular Technology}, 2026.}
}

\begin{document}

\makeatletter
\def\ps@IEEEtitlepagestyle{
  \def\@oddfoot{ \hfill}
  \def\@evenfoot{}
}

\title{Unique Word Channel Estimation for Oversampled~OTFS}

\author{Radim Zedka,~Roman Marsalek,~\IEEEmembership{Member,~IEEE,} Marek Bobula, and~Arman~Farhang~\IEEEmembership{Senior Member,~IEEE}
             \thanks{
            \copyright 2026 IEEE. Personal use of this material is permitted. Permission from IEEE must be obtained for all other uses, in any current or future media, including reprinting/republishing this material for advertising or promotional purposes, creating new collective works, for resale or redistribution to servers or lists, or reuse of any copyrighted component of this work in other works.
            Roman Marsalek is with the Department of Radio Electronics, Brno University of Technology, 616 00 Brno, Czech Republic (e-mail: marsaler@vut.cz).
            Marek Bobula is with RACOM, s.r.o., 592 31, Nove Mesto na Morave, Czech Republic (e-mail: marek.bobula@racom.eu).
            Radim Zedka is with both of the above institutions (e-mail: 164441@vut.cz, radim.zedka@racom.eu).
            Arman Farhang is with the Department of Electronic and Electrical Engineering, Trinity College Dublin, Dublin 2, D02 PN40 Ireland (e-mail: arman.farhang@tcd.ie).
            
            This publication has emanated from research jointly funded by Ministry of Education, Youth and Sports of the Czech Republic under the project LUC24141 \emph{Joint Communication and Sensing to Enhance Robustness of 6G Systems} (program INTER-EXCELLENCE II), and by Taighde \'Eireann - Research Ireland under Grant number 21/US/3757.  
            }
        }
\maketitle

\thispagestyle{specialfooter}

\begin{abstract}
 Practical aspects of orthogonal time frequency space (OTFS), such as channel estimation and its performance in fractional delay-Doppler (DD) channels, are a lively topic in the OTFS community.
 Oversampling and pulse shaping are also discussed in the existing literature, but not in the context of channel estimation. 
 To the best of our knowledge, this paper is the first to address the problem of data-to-pilot and vice versa energy leakage caused by oversampling and pulse shaping in OTFS. Theoretical analysis is performed on an oversampled, pulse-shaped OTFS implementing the embedded pilot channel estimation technique, revealing a trade-off between the amount of energy leakage and excess bandwidth introduced by the pulse shape.
 Next, a novel variant of OTFS is introduced, called UW-OTFS, which is designed to overcome the leakage problem by placing the pilot in the oversampled time domain instead of the DD domain. 
 The unique structure of UW-OTFS offers 36 percent higher spectral efficiency than the OTFS with embedded pilot. UW-OTFS also outperforms traditional OTFS in terms of bit error ratio and out-of-band emissions.  
\end{abstract}


\section{Introduction}

Achieving reliable wireless communication in high-speed scenarios, such as ground-to-air links or bullet train networks, is a great challenge. The main difficulty arises from the rapidly changing multipath channel, where traditional orthogonal frequency division multiplexing (OFDM) struggles to handle extreme time variability \cite{b_OTFS_OFDM_1}. 
To address this issue, a new delay-Doppler (DD) domain multiplexing scheme called orthogonal time frequency space (OTFS) was developed in \cite{b_OTFS_orig}.
Unlike OFDM, which was developed for frequency-selective channels with low mobility, OTFS is designed to cope with the doubly-dispersive channel effects by introducing an additional precoding known as the inverse symplectic Fourier transform (ISFT). 
Using this precoding operation, OTFS spreads the input data symbols in the frequency and time domains, which, in the case of ideal transmit/receive (Tx/Rx) pulse shaping can harvest full diversity of the channel \cite{b_OTFS_diversity}. 

Recently, there has been an increasing focus in the research community on the practical aspects of OTFS, such as channel estimation \cite{b_OTFS_OFDM_1, b_Pfadler_GLOBECOM, b_OTFS_SpectEff_IoT_2023, b_Raviteja_TVT, b_Hashimoto_ICC, b_TwoStage_CE_OTFS} and pulse shaping (preceded by oversampling) \cite{b_FFT_FBMC, b_DFTs_OTFS, b_OTFS_spect_shape, b_oversampled_2024}. 
If we compare the channel estimation process in OTFS with that in OFDM, we find several major differences.
OFDM estimates most of the channel state information (CSI) using dedicated pilot symbols, typically transmitted before the payload. Additionally, to compensate for the carrier frequency offset and small time variations of the CSI, OFDM replaces some subcarriers in the payload with known pilots. 
This approach works perfectly with oversampling and pulse shaping.
OTFS, on the other hand, requires much more sophisticated CSI estimation methods, and oversampling and pulse shaping are not considered in the existing literature. 
An example of channel estimation in OTFS may be the embedded (impulse) pilot method \cite{b_Raviteja_TVT, b_OTFS_OFDM_1, b_OTFS_SpectEff_IoT_2023}, where a portion of the DD domain data matrix is reserved for a pilot surrounded by guard (zero) symbols. 
Once the received signal is transformed back to the DD domain, the CSI is estimated by extracting the pilot and its surrounding region in the DD domain. 
A different method is presented in \cite{b_TFdomain_OTFS_CE}, where the pilots are located in the frequency-time (FT) domain.
In this work, the OTFS is designed to cope with fractional delays, using the multiple signal classification (MUSIC) algorithm, and with fractional Doppler components, using a line fitting. 
However, prior knowledge of the channel tap count of the DD domain is necessary for the Rx side, which is impractical.
In \cite{b_Superimposed_pilot_OTFS}, the pilots are superimposed directly onto the DD domain data matrix. However, in this work, prior knowledge of the channel tap count is also required on the Rx side.
This requirement is removed in \cite{b_low_overhead_OTFS}, where pilots are also located in the FT domain. However, their system was designed for integer Doppler channels, which do not appear in practice since a fractional Doppler always exists.
In \cite{b_Fettweis_TWC}, an OTFS-like system is designed with Zadoff-Chu sequences at the beginning and end of the transmission, each protected by a cyclic prefix (CP). However, as also admitted in \cite{b_Fettweis_TWC}, the Zadoff-Chu sequence is not optimal for doubly dispersive channel estimation.
In a more recent work \cite{b_Chockalingam_CL2024} an iterative CSI estimation algorithm for the Zak transform OTFS is introduced, which is capable of estimating fractional delay and Doppler components. 
However, this work uses single-stage implementation of OTFS with linear pulse shaping while our work focuses on two-stage OTFS with circular pulse shaping performed by the OFDM modulator.
In \cite{b_TimeDom_CE}, a study on OTFS in fractional Doppler channels and residual synchronization errors is presented. 
It introduces the idea of a time-domain estimation of the doubly dispersive channel with a fractional Doppler by using spline interpolation to find the sample-by-sample channel estimate. 
 In \cite{b_DiffMod_OTFS}, differential modulation is implemented in OTFS and decision feedback approach assisted by lightweight neural network ensures channel estimation and data detection. This design offers significant savings in resources because it does not use any pilot.

In all of the above-mentioned works discussing channel estimation (except \cite{b_Chockalingam_CL2024}), oversampling and pulse shaping are not considered. 
However, for practical OTFS waveforms, oversampling and pulse shaping are necessary to prepare the transmit signal to comply with the spectral mask and reduce the length of the transmit filter which contributes to the overall channel length. 
However, oversampling and pulse shaping often cannot be simply introduced to the existing OTFS channel estimation methods that we mentioned above, as it may cause severe performance degradation. 
For example, if oversampling and pulse shaping are introduced to the embedded pilot channel estimation method in \cite{b_Raviteja_TVT}, convolution with the pulse shaping filter impulse response in the delay dimension is inevitable, causing the data energy to leak into the pilot and vice versa. 
This energy leakage then degrades the quality of the CSI estimate and causes a severe error floor even for a high signal-to-noise ratio (SNR) as shown in the simulations section of this paper. 
Although leakage can be reduced by adjusting the roll-off factor of the pulse-shaping filter, this comes at the cost of increased excess bandwidth.

Due to the lack of literature on CSI estimation in pulse-shaped oversampled OTFS, this paper starts by performing a leakage analysis on a system called CP-OTFS. That can be described as an embedded pilot OTFS with oversampling, pulse shaping, and a cyclic prefix attached to each delay block, hence the CP in CP-OTFS.
Although the name CP-OTFS has been used in several publications, our version of CP-OTFS was mainly based on the following ideas: \cite{b_Raviteja_TVT} - the embedded impulse pilot; \cite{b_TimeDom_CE} - multiple CP design and separate delay block Rx processing; \cite{bayat2023unified} - oversampling and pulse shaping. 

\begin{figure}[t] \centering
\includegraphics[scale=0.30]{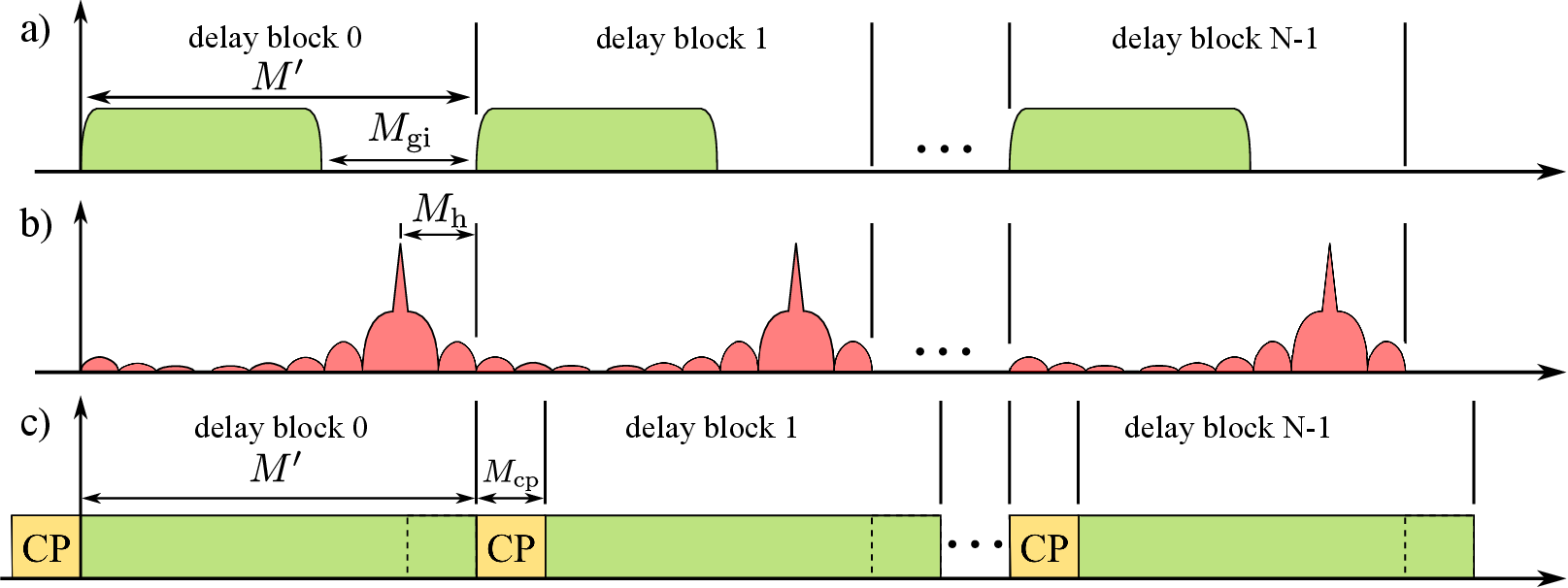}\\
\caption{Waveform diagrams of: a) the data part of UW-OTFS, b) the UW pilot of UW-OTFS, c) CP-OTFS.}
\label{fig_UW_OTFS_waveforms}
\end{figure}

In response to the leakage problem of CP-OTFS, we offer an innovative solution by presenting a new OTFS system, called the unique word OTFS (UW-OTFS). 
UW-OTFS estimates CSI in the oversampled time domain, therefore, completely avoiding the data-to-pilot interference issue. 
To make this possible, UW-OTFS uses the popular UW-OFDM \cite{b_Huemer_UWODFM_1, b_Nonsys_OFDM}, generating a guard interval (GI) within each delay block of the Tx waveform - see waveform a) in Fig.~\ref{fig_UW_OTFS_waveforms}. 
Due to this feature, UW-OFDM has been adopted in several multicarrier schemes reported in the literature. 
In \cite{b_Sahin_UW}, a discrete Fourier Transform (DFT)-spread OFDM uses UW-OFDM for better tail suppression, however, time-invariant channel with perfect CSI knowlegde is assumed. 
Another adaptation of UW-OFDM (with the same assumptions) can be found in \cite{b_UW_GDFM}, where UW-OFDM from \cite{b_Huemer_UWODFM_1} is used in a generalized frequency domain multiplexing (GFDM) scheme, offering better performance than UW-OFDM and CP-OFDM. 
In \cite{b_Ehsanfar2020_trans}, UW-GFDM is proposed for block-fading channels, elaborating on the need to have GI included within the FFT block.
In our design, GI serves two purposes: 
\begin{enumerate}
    \item It allows separate delay block processing, similar to multiple-CP OTFS, thus avoiding large matrix inversions required in single-CP OTFS designs.

    \item It creates space for UW pilot placement, which is used for CSI estimation in the oversampled time domain - see waveform b) in Fig.~\ref{fig_UW_OTFS_waveforms}.    
\end{enumerate}  
As the name already suggests, UW-OTFS uses a UW pilot, representing an alternative to the widely used impulse pilot \cite{b_Raviteja_TVT} embedded in a DD grid and used for the estimation of CSI in most current OTFS implementations. 
Thanks to GI, the UW pilot can be extracted on the Rx input and used for CSI estimation, immune to data-to-pilot energy leakage.

Although the UW pilot was originally contained within the GI \cite{b_Huemer_UWODFM_1, b_Nonsys_OFDM}, we expand the UW pilot outside the GI, throughout the delay block, allowing us to optimize it for better power spectral density (PSD) properties and, most of all, reduce the out-of-band (OOB) emissions of the oversampled waveform.

The ability to separately process delay blocks at the receiver is natural for both CP-OTFS and UW-OTFS. 
They both belong to the group of OFDM-based OTFS systems (a.k.a. two-stage OTFS or circularly pulse-shaped OTFS) \cite{b_TimeDom_CE, b_CP_OTFS_WCL, b_Novel_CP_OTFS_2023, b_Farhang_ICC} which typically use a zero guard interval or a CP before each delay block (see waveform c) in Fig.~\ref{fig_UW_OTFS_waveforms}), and, therefore, offer low receiver complexity by avoiding large matrix inversions or iterative algorithms that are typically needed for single-CP or no-CP OTFS designs \cite{b_Raviteja_Practical_pulse, b_Viterbo_TWC, b_Raviteja_TVT, b_OTFS_Rapid_Multipath, b_OTFS_OFDM_1, b_Pfadler_GLOBECOM, b_LMMSE_low_Complex_OTFS, b_Superimposed_pilot_OTFS, b_low_overhead_OTFS}.

Furthermore, to keep receiver processing linear and its complexity low in both UW-OTFS and CP-OTFS, we use the generalized complex exponential basis expansion model (GCE-BEM) \cite{b_Farhang_BEM_ICC, b_BEM_Gonzales_WCL2024, b_BEM_Turbo_TCOMM2024, b_BEM_TCOMM2022} and the linear minimum mean square error (LMMSE) approach for CSI estimation, as well as for data estimation.

To the best of our knowledge, no study has been published that discusses CSI estimation in two-stage OTFS with oversampling and pulse shaping. In addition, there is no other prior work that effectively combats the data-to-pilot leakage as the UW-OTFS that we introduced in this paper.
The contributions of this paper are summarized as follows:
\begin{itemize}
    \item To highlight the energy leakage issue, we provide a detailed leakage analysis of the CSI estimation process in CP-OTFS. 
    A tradeoff between excess bandwidth and the amount of data-to-pilot energy leakage is revealed.
    
    \item As a solution to the leakage issue, we propose a UW-OTFS modem that outperforms CP-OTFS in terms of bit error ratio (BER) while also achieving lower spectral OOB emissions and $36 \%$ higher spectral efficiency, based on our numerical results.
    
    \item To allow a fair comparison, we develop a linear CSI estimation technique based on GCE-BEM and LMMSE for both UW-OTFS and CP-OTFS.

    \item We analyze the computational complexity of UW-OTFS and CP-OTFS and evaluate it in terms of the number of complex multiplication equivalents and memory footprint.
    
\end{itemize}

The remainder of the paper is structured as follows. Section~II presents OTFS principled in continuous time, and Section~III introduces the discrete-time CP-OTFS with oversampling and pulse shaping. 
Section~IV analyzes the delay dimension energy leakage in CP-OTFS. In Section~V, a novel system is introduced, denoted UW-OTFS, and Section~VI presents the novel channel estimation method used in UW-OTFS. 
In Section~VII, the same channel estimation method is implemented in CP-OTFS. Section~VIII contains the computational complexity analysis of both systems. Section~IX provides numerical analysis of the achievable spectral efficiency and a numerical comparison of the computational complexity and memory footprint. 
Section~X contains the results of the Monte Carlo simulations, and Section~XI concludes the paper.

\subsection{Notation}
We denote $\sigma_w^2$ as the additive white Gaussian noise (AWGN) variance. 
The m-ary quadrature amplitude modulation (QAM) symbols are always normalized to unit variance. 
 $\E\{\cdot\}$, $\tr\{\cdot\}$, $(\cdot)^*$, $(\cdot)^{\rm T}$, $(\cdot)^{\rm H}$, $(\cdot)_M$, $\floor{\cdot}$, $\ceil{\cdot}$ and $\nint{\cdot}$ represent the expectation operator, trace operator, complex conjugation, matrix transpose, conjugate matrix transpose, modulo-$M$ operation, floor, ceil, and round operations, respectively. 
Infinite discrete signals and sequences are defined by normal lowercase letters with square brackets, e.g., $s[k]$ $\forall$ $k \in \mathbb{Z}$.
Column vectors are defined by lowecase bold letters, e.g., $\mathbf{h}$, and the matrices are defined by uppercase bold letters, e.g., $\mathbf{X}$, where the $M \times M$ unit matrix is denoted as $\mathbf{I}_M$. 
Multi-dimensional matrices (tensors) are defined as uppercase bold letters with a bar, e.g. $\mathbf{\bar{X}}$.
Addressing the entries of any vector/matrix/tensor is done with square brackets, e.g., the $p$-th row and $q$-th column of a matrix $\mathbf{X}$ is expressed as $X[p,q]$, where the indexing starts with $0$. 
A similar method is also used to address sets of numbers, e.g., the $0$-th element of the set $\mathcal{M} \coloneqq \{3,4,5,6\}$ is denoted as $\mathcal{M}[0] = 3$. 

To denote closed and open intervals, we use $[ \cdot, \cdot ]$ and $(\cdot, \cdot)$, respectively.

The discrete Dirac delta impulse is defined as $\delta[k] \ \forall \ k \in \mathbb{Z}$, where $\delta[0]=1$ and zero otherwise, while $\delta(t)$ represents the continuous Dirac delta function. 
$\mathbf{0}_{M \times N}$ is an all-zeros matrix of size $M \times N$.
The symbol $\tau'$ denotes the oversampled delay domain. 

\section{Basic Definition of OTFS}

Considering a delay-Doppler grid of size $M \times N$, where $M$ is the number of delay bins, and $N$ the number of Doppler bins, we define an $M \times N$ matrix of m-ary quadrature amplitude modulated (m-QAM) data symbols $\mathbf{D}$.
It is then transformed into the FT domain using the ISFT, $ \mathbf{X} = \mathbf{F}_{M} \mathbf{D} \mathbf{F}_N^{\rm H}$, 
where $\mathbf{F}_{M}$ and $\mathbf{F}_N$ are the discrete Fourier transform (DFT) matrices of size $M \times M$ and $N \times N$, respectively, with 
$F_{M} [p,q] = 1/\sqrt{M} e^{-j2\pi\frac{pq}{M}}$ for $p,q \in [ 0, M-1]$.

The FT domain matrix is then transformed into a base-band time domain signal $s(t)$ using the Heisenberg transform, defined as \cite{b_Viterbo_TWC}
\begin{equation}\label{eq_Heisenberg_def}
  s(t) = \sum_{m = 0}^{M-1} \sum_{n = 0}^{N-1} X[m,n] g_{\rm tx}(t - nT) e^{j 2 \pi m \Delta f (t- nT) },
\end{equation}
where $T$ is a time spacing between the columns of $\mathbf{X}$, $\Delta f = 1/T$ is a frequency spacing between the rows of $\mathbf{X}$ and $g_{\rm tx}(\cdot)$ is a time domain window function (different from the pulse shaping filter, which will be introduced in the next section). 
OTFS signal then passes through the linear time-variant (LTV) channel, yielding a time-domain Rx signal
\begin{equation}\label{eq_r_t_cont_def}
  r(t) = \iint h(\tau, \nu) s(t - \tau) e^{j 2\pi \nu (t- \tau)}  d \tau  d \nu.
\end{equation}
Here $h(\cdot, \cdot)$ is a DD domain channel function, typically assumed to be a sparse model
\begin{equation}\label{eq_LTV_cont_def}
   h(\tau, \nu) = \sum_{i = 0}^{P-1} h_i \delta(\tau - \tau_i) \delta(\nu - \nu_i),
\end{equation}
where $P$ is the channel tap count, $h_i$ is the $i$th channel tap complex gain, $\delta(\cdot)$ is the Dirac delta function, $\nu_i$ is the $i$th Doppler tap position from within range $[ -\nu_{\rm max}, \nu_{\rm max})$ with $\nu_{\rm max}$ being a maximum Doppler spread.
The delay tap $\tau_i$ is typically a positive real number.

On the Rx side, OTFS performs matched filtering using the Wigner transform, which yields the $M \times N$ FT domain Rx matrix $\mathbf{Y}$, defined as
\begin{equation}\label{eq_Y_cont_def}
   Y[m',n'] = \int g_{\rm rx}^*(t - n'T) r(t) e^{-j 2\pi m' \Delta f (t- n'T)}  d t,
\end{equation}
and where $g_{\rm rx}(\cdot)$ is the Rx time domain window. 
The continuous-time model of OTFS presented above cannot be directly implemented into hardware because, in reality, every communication waveform requires oversampling and pulse shaping. A more practical discrete-time OTFS model will be discussed in the next section.

\section{OTFS with Pulse Shaping} 
Pulse-shaping (preceded by oversampling) is a necessary part of every digital communication system. Its main role is to prepare the information signal for transmission over the wireless channel with limited bandwidth \cite{farhang2010signal}.
As mentioned earlier, the pulse-shaping filter effects at the transmitter/receiver are not taken into account in the existing literature on channel estimation of OTFS \cite{b_Raviteja_TVT, b_OTFS_OFDM_1,b_Hashimoto_ICC, b_TwoStage_CE_OTFS, b_low_overhead_OTFS}. 
Below, we define the transmitter, the discrete-time LTV channel, and the receiver of the CP-OTFS, which is an oversampled version of the embedded pilot OTFS we developed based on \cite{b_Raviteja_TVT, b_TimeDom_CE, bayat2023unified}.

\subsection{CP-OTFS Transmitter}

First, we introduce oversampling and pulse shaping to the matrix in the FT domain, $\mathbf{X}$.
In OFDM, oversampling and pulse shaping is typically performed by inserting a number of inactive subcarriers next to the data subcarriers \cite{farhang2010signal}. 
Alternatively, oversampling can be achieved by first cyclically extending the frequency dimension of $\mathbf{X}$ from $M$ to $M' = QM$, where $Q \in \mathbb{N}$ is the oversampling factor.
For this, we define an $M' \times M$ cyclic extension matrix $\mathbf{A}_{\rm d}$ with elements $A_{\rm d}[p,q] = \delta\big[(p-q)_M \big]$. 
After oversampling, we apply circular pulse shaping, by multiplication with a pulse shape vector $\boldsymbol{\psi}$ of size $M' \times 1$, which is the frequency response of the root raised cosine (RRC) filter with a roll-off factor $\alpha$ \cite{bayat2023unified}. 
The elements of $\psi[p]$ are given by:
\begin{equation}\label{eq_RRC_def}
\begin{cases}
\varphi(p - M/2) \quad \forall \quad p \in [ -M_\alpha, M_\alpha - 1] + M/2\\
1 \quad \forall \quad p \in [ 0, M/2 - M_\alpha -1  ] \ \cup \ [ M_\alpha -M/2, -1  ]+M'\\
\varphi(M'-M/2 - p) \quad \forall \quad p \in [ -M_\alpha, M_\alpha - 1] + M'-M/2\\
0 \quad \text{otherwise},
\end{cases}
\end{equation}
where, $\varphi(p) = \sqrt{2+2\cos \big(\pi (p+M_\alpha)/(2M_\alpha) \big)}$, and $M_\alpha = \floor{\alpha M/2}$ is the number of excess subcarriers, causing an increase of the total active subcarrier count to $M_{\rm s} = M + 2M_\alpha$. 
After applying the Heisenberg transform - performing inverse DFT (IDFT) of length $M'$ across the frequency domain (FD) - and CP addition, we obtain an $M_{\rm x}' \times N$ transmit signal matrix, defined as $\mathbf{S}_{\rm cp} = \mathbf{A}_{\rm cp}\mathbf{F}_{M'}^* \mathbf{\Psi} \mathbf{A}_{\rm d} \mathbf{X}$. 
Here, $M_{\rm x}' = M' +M_{\rm cp}$, the $M_{\rm x}' \times M'$ CP addition matrix is defined as $A_{\rm cp}[p,q] = \delta\big[(p-q-M_{\rm cp})_{M'} \big]$ and $\Psi [p,q] = \delta[p-q]\psi[p]$. 
The matrix $\mathbf{S}_{\rm cp}$ can be expressed as an oversampled time-domain signal $s[k] = S_{\rm cp} \big[(k)_{M_{\rm x}'}, \floor{k/M_{\rm x}'} \big]$, for $k \in [ 0, M_{\rm x}'N-1 ]$ and zero otherwise, so the elements of $s[k]$ can be identically written as
\begin{equation}\label{eq_mCP_TX_s_k_PS_def}
    \sum_{m'=0}^{M'-1} \psi[m'] \sum_{n=0}^{N-1} X[(m')_{M},n] g_{\rm tx}[k-nM_{\rm x}'] e^{j2\pi \frac{m'(k-nM_{\rm cp})}{M'}}.
\end{equation}
Here, the Tx discrete windowing function for the CP-OTFS is defined as $g_{\rm tx}[k] = 1/\sqrt{M}$ for all $k \in [ -M_{\rm cp}, M'-1 ]$ and zero otherwise.
The CP-OTFS signal consists of $N$ delay blocks, each $M'$ samples long with an attached CP of length $M_{\rm cp}$ \cite{b_CP_OTFS_WCL}. 
Note, that to ensure zero interference between the consecutive delay blocks, the CP length must satisfy 
\begin{equation}\label{eq_CP_Channel_Memory}
M_{\rm cp} \geq L'-1 ,
\end{equation}
where $L'$ is the oversampled channel length.

\subsection{Linear Time-Variant Channel Model}

We consider a sparse LTV channel, with $P$ paths, delay shifts $k_{{\tau'}}[i]$, Doppler shifts $\xi[ i]$ and complex gains $h[i]$, for paths $i\in \{0,\ldots,P-1\}$. 
The three CSI parameters, $h[i]$, $k_{{\tau'}}[i]$, and $\xi[ i]$ are considered mutually independent random variables. 
The oversampled OTFS transmit signal $s[\cdot]$  propagated through the LTV channel yields
\begin{equation}\label{eq_yk_channel_def}
r[k] = \sum_{i = 0}^{P -1}  h[i] s [k-k_{{\tau'}}[i] ]    e^{j\frac{2\pi \xi[ i] k }{M'}} + w[k],
\end{equation}
where $w[\cdot]$ represents AWGN with zero mean and the variance $\sigma_{ w}^2$, i.e., $w[\cdot]\sim \mathcal{CN}(0,\sigma_{ w}^2)$. 
Since we focus on OFDM-based implementation of OTFS, we consider Doppler shifts $\xi[i]$ with both integer and fractional parts being normalized to the subcarrier spacing of OFDM, i.e., $\Delta f=\frac{F}{M'}$, where $M'$ represents the number of delay bins and $F$ is the sampling frequency which depends on the transmit signal bandwidth. 
The Doppler shifts $\xi[i]$ lie within the range $[ -\nu_{\rm max}, \nu_{\rm max})/\Delta f$, where $\nu_{\rm max}$ is the maximal expected Doppler frequency,
calculated as $\nu_{\rm max} = \mathscr{v}_{\rm max} f_0 /\rm c_0$ and where, $\mathscr{v}_{\rm max}$ is the maximum expected velocity in $ \rm km/h$, $f_0$ is the carrier frequency in $\rm Hz$ and $\text{c}_0$ is the speed of light in $\rm km/h$.

As the sampling frequency in wideband systems is very high and \eqref{eq_yk_channel_def} represents the oversampled signal, it is reasonable to approximate delay shifts with integer values \cite{b_OTSM_2021}.
Hence, delay shifts $k_{{\tau'}}[i]$ can take integer values from $0$ to $L'-1$ where $L'$ is the oversampled channel length, defined as $L' = \nint{\tau_m F} +1$, where $\tau_m$ is the maximum channel delay spread.

\subsection{CP-OTFS - Basic Receiver Processing}

The Rx input signal $r[\cdot]$ is processed with a discrete Rx windowing function, defined as
\begin{equation}\label{eq_g0_window_def}
g_{0}[k] = 1/\sqrt{M'} \quad \forall \quad k \in [ 0, M'-1 ],
\end{equation}
 and zero otherwise, which discards the CP.
The FT domain $M' \times N$ matrix $\mathbf{Y}$ is defined, using the Wigner transform, as
\begin{equation}\label{eq_mCP_Y_mat_def}
Y[m,n] =  \sum_{k' = -\infty}^{\infty} g_{0}[ k'-n M_x' ] r[k'] e^{-j2\pi \frac{m(k'- nM_{\rm cp})}{ M'}},
\end{equation}
which, as long as \eqref{eq_CP_Channel_Memory} holds, can also be written as (AWGN neglected for simplicity)
\begin{equation}\label{eq_mCP_Y_mat_eval_4}
Y[m,n] =  \sum_{m'=0}^{M-1} H_n[m, m'] X[m',n],  
\end{equation}
where the $n$-th equivalent channel matrix (ECM), $\mathbf{H}_n$, of size $M' \times M$ is defined as
\begin{equation}\notag
H_n[m, m'] =   \sqrt{Q}  \sum_{i = 0}^{P -1}  h[i]   e^{j2\pi \xi[ i] \frac{  nM_x' }{M'}} e^{-j2\pi  \frac{ k_{\tau'}[i] m'}{M'} } 
\end{equation}
\begin{equation}\label{eq_mCP_ECT_PS_def}
\times \sum_{k = 0}^{Q-1} e^{-j2\pi  \frac{ k_{\tau'}[i] k}{Q} } \psi[m'+kM]  \chi \big(m'+kM - m + \xi[ i], M'\big).
\end{equation}
Here, $\chi(\cdot, \cdot)$ is the Dirichlet kernel, defined for all $z \in \mathbb{R}$ and $n\in \mathbb{N}$ as
\begin{equation}\label{eq_mCP_Dirichlet_kern_def}
\chi(z, n) = \frac{1}{n}\sum_{k = 0}^{n-1}     e^{j2\pi  \frac{ k z}{n} }.
\end{equation}
The CSI estimation part of the CP-OTFS receiver processing will be elaborated as part of the leakage analysis in the next section.

\section{Delay Leakage Analysis}

 The interaction of the oversampled integer delay LTV channel in \eqref{eq_yk_channel_def} with the transmit/receive pulse-shaping filters and downsampling leads to fractional delay shifts. 
This causes serious leakage issues along the delay dimension.

Hence, in this section, we embed the impulse pilot to the CP-OTFS and then we analyze the symbol energy leakage issues along the delay dimension, considering the transmit and receive pulse-shaping filters.

\subsection{Embedded Impulse Pilot}
The impulse pilot is embedded to the DD domain matrix $\mathbf{D}$, where a single entry $D[p,q]$ is now defined as
\begin{equation}\label{eq_CP_x_DD_mat_def}
 \begin{cases}
\rho_0 \sqrt{M_0} \quad \forall \quad p = p_0 \ \cap \ q = q_0
\\
0 \quad \forall \quad p \in [ 0, M_{\rm g}-1 ] \setminus \{p_0\} \cap \ q \in [ 0,N-1] \setminus \{q_0\}\\
\text{QAM data symbols} \quad \text{otherwise},
\end{cases}
\end{equation}
where $p_0 = \floor{(M_{\rm g}-1)/2}$ and $q_0 = N/2-1$ are the delay and Doppler pilot position indices, respectively, $M_{\rm g}$ is the pilot guard size along the delay dimension, $M_0 = M_{\rm g} N$ is the total number of delay-Doppler bins belonging to the pilot guard region and $\rho_0^2$ is the pilot energy coefficient.
Since we assume a realistic LTV channel model with fractional Doppler taps, we use the full Doppler guard, covering all $N$ taps of the Doppler domain with zeros \cite{b_Raviteja_TVT, b_OTFS_SpectEff_IoT_2023}. 
Typically, the embedded pilot CSI estimation requires bringing the received signal into the DD domain \cite{b_Raviteja_TVT, b_Pfadler_GLOBECOM}.
Therefore, we will now analyze the behavior of a single input data symbol $D[p,q] \equiv x_0$, located at $p \equiv p_0$ and $q \equiv q_0$, in the received DD domain matrix, after propagating through the oversampled LTV channel, defined in \eqref{eq_yk_channel_def}.

\subsection{Leakage Analysis}
We begin by replacing $X [m',n]$ in \eqref{eq_mCP_Y_mat_eval_4} with
\begin{equation}\label{eq_X_FT_mat_def}
 X[m',n] = x_0/\sqrt{MN} 
    e^{-j2\pi \frac{m' p_0}{M}} e^{j2\pi \frac{nq_0}{N}}.
\end{equation}
For greater clarity of the analysis, we will now assume the channel has only a single tap (i.e., $P=1$, $\xi[ i] \equiv \xi_0$, $k_{\tau'}[i] \equiv k_{\tau'0}$ and $h[i] \equiv h_0$) and that it has zero Doppler component, i.e., $\xi_0 = 0$.
Due to the zero Doppler component, the Dirichlet kernel $\chi(z, M')$ in \eqref{eq_mCP_ECT_PS_def} collapses into a discrete Dirac delta function $\delta\big[ (k)_{M'} \big]$ for all $k \in \mathbb{Z}$.

At this point, we must reduce the dimension of $\mathbf{Y}$ in \eqref{eq_mCP_Y_mat_eval_4} from $M' \times N$ to $M \times N$, otherwise, symplectic Fourier transform (SFT) could not be used to obtain the $M \times N$ DD domain matrix $\mathbf{\hat{Y}}$ necessary for CSI estimation.
The Rx pulse shaping and downsampling of $\mathbf{Y}$ is achieved by element-wise multiplication with $\mathbf{\psi}$ (i.e., matched filtering) followed by aliasing, which is performed by multiplication with $\mathbf{A}_{\rm d}^{\rm T}$ of size $M \times M'$. 
Therefore, the decimated FT matrix of size $M \times N$ is given by $\mathbf{\hat{Y}} = \mathbf{A}_{\rm d}^{\rm T} \boldsymbol{\Psi} \mathbf{Y}$. 
Its elements are defined as
\begin{equation}\label{mCP_Y_hat_mat_final}
\hat{Y}[m,n]  =   \frac{x_0 h_0 \sqrt{Q}}{\sqrt{MN}}    e^{j2\pi \frac{n q_0}{N} }   e^{-j2\pi \frac{m (p_0+ k_{\tau'}/Q)}{M}} \hat{\psi}[m],
\end{equation}
where $\boldsymbol{\hat{\psi}}$ is a vector of size $M \times 1$, with elements
\begin{equation}\label{mCP_Xi_vec_def}
 \hat{\psi}[m] = \sum_{k=0}^{Q-1}\psi^2[m+kM]e^{-j2\pi \frac{  k_{\tau'}k}{Q}} . 
\end{equation}
The $M \times N$ DD domain matrix $\mathbf{\hat{D}}$ is given by the SFT,
\begin{equation}\label{eq_mCP_y_DD_mat_eval}
\hat{D}[m,n] =  \frac{1}{\sqrt{MN}} \sum_{l=0}^{M-1} \sum_{k=0}^{N-1} 
    \hat{Y}[l,k] e^{j2\pi\big( \frac{ml}{M} - \frac{nk}{N}\big)},
\end{equation}
which, after some more manipulations, yields the final formula for the DD domain receiver matrix
\begin{equation}\label{eq_mCP_y_DD_mat_final}
\hat{D}[m,n]  =  x_0 h_0 \sqrt{Q}  \delta [ n-q_0 ] \phi[m],
\end{equation}
where $\boldsymbol{\phi}$ is an $M \times 1$ vector, defined as
\begin{equation}\label{eq_mCP_Phi_vec_def}
\phi[m] = \frac{1}{M} \sum_{l=0}^{M-1} e^{-j2\pi\frac{l(p_0 -m + k_{\tau'}/Q)}{M} }  \hat{\psi}[l].
\end{equation}
The presence of $\delta [ n-q_0 ]$ in \eqref{eq_mCP_y_DD_mat_final} indicates that the only non-zero column of $\mathbf{\hat{D}}$ has the index $n=q_0$, due to the zero Doppler component of the channel. 
From \eqref{mCP_Xi_vec_def} it is clear that if $k_{\tau'} = Qk$ for $k \in \{0,1, 2, \dots\}$ (i.e., in the case of integer downsampled delay channel taps), we get $\hat{\psi}[m] = 1$ for all $m \in [ 0, M-1]$, which is the Nyquist property of the raised cosine filter. 
Therefore, the DD domain matrix yields
\begin{equation}\label{eq_mCP_y_DD_mat_final_even}
\hat{D}[m,n]  =   x_0 h_0 \sqrt{Q}  \delta \big[ (n-q_0)_N \big] \delta \big[ (p_0 -m + k)_{M}\big],
\end{equation}
proving that all the energy of $x_0$ is localized within a single bin of $\hat{D}[m,n]$ with a column index $n=q_0$ and row index $m$, satisfying $(p_0 -m + k)_{M} = 0$. A detailed derivation of the above steps can be found in Appendix A.

The situation becomes complicated for fractional delays, i.e., if $k_{\tau'} \neq Qk$ for $k \in \{0,1, 2, \dots\}$, because here the Nyquist property no longer holds. 
Formula \eqref{eq_mCP_Phi_vec_def} cannot be simplified into a Dirac impulse, therefore, the energy of $x_0$ is dispersed along the delay dimension of $\mathbf{\hat{D}}$. 

\begin{figure}[t] \centering
\begin{tabular}{c} 
\includegraphics[scale=0.50]{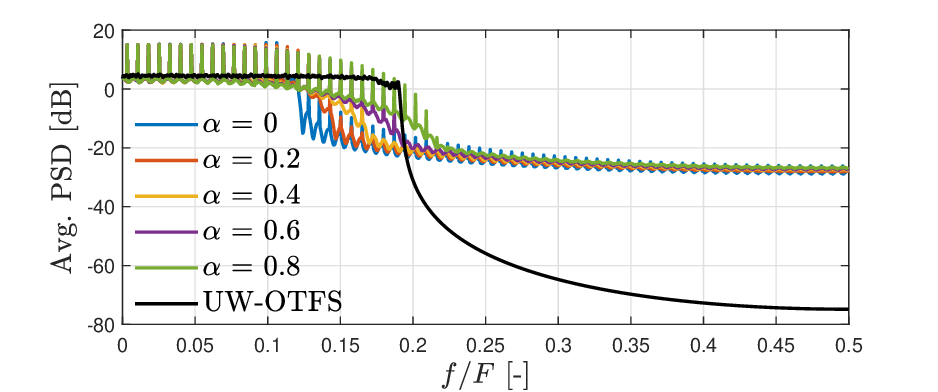}\\ \includegraphics[scale=0.50]{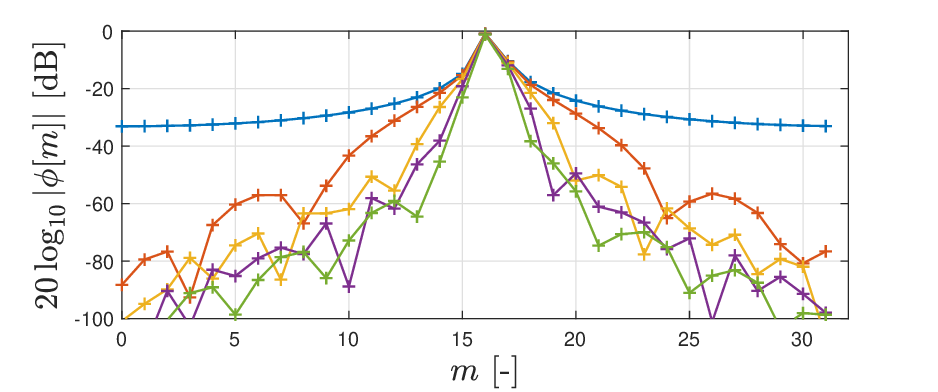}
\end{tabular}
\caption{Embedded pilot PSD comparison (top) and energy dispersion in the DD domain (bottom) for various values of the roll-off factor.}
\label{fig_EP_sym_energy_DD}
\end{figure}

The lower part of Fig.~\ref{fig_EP_sym_energy_DD} illustrates the case for $M = 32$, $Q=4$, $p_0 = 16$, $k_{\tau'} = 1$ and $\alpha = \{0, 0.2, 0.4, 0.6, 0.8\}$. 
From this we can quickly estimate the interference caused by one complex symbol in neighboring delay bins. 
For example, if $\alpha = 0$, there will be $> 1 \%$ of the symbol energy present in each delay bin (that is, the leakage energy is $\geq -40$ dB).
Moreover, if we consider energy $ > 1 \%$ as a notable interference, the total number of delay bins affected by this interference will be $ \{31, 11, 6, 4, 3\}$ for $\alpha = \{0, 0.2, 0.4, 0.6, 0.8\}$.
Therefore, an increase of $\alpha$ decreases the number of delay guard bins $M_{\rm g}$ required to prevent the interference. 
On the other hand, $\alpha$ is responsible for generating the excess subcarriers according to $M_\alpha = \floor{\alpha M/2}$, and this influence is depicted in the top part of Fig.~\ref{fig_EP_sym_energy_DD} (the last curve represents the proposed UW-OTFS that will be introduced in the following section).

\begin{figure}[t] \centering
\includegraphics[scale=0.7]{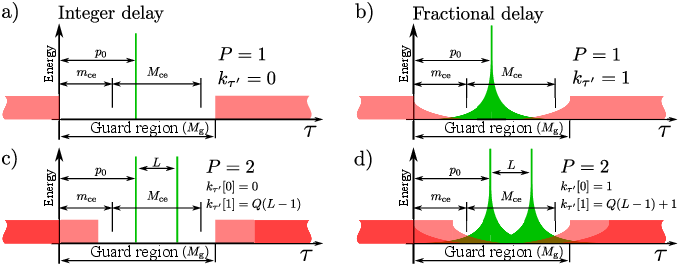} 
\caption{Illustration of the delay dimension leakage in the energy of $\mathbf{\hat{D}}$.} 
\label{fig_DD_leakage}
\end{figure}

To provide additional insight into the energy leakage issue, a simple visualisation of the delay domain of $\mathbf{\hat{D}}$ is shown in Fig.~\ref{fig_DD_leakage}.
Here, we see the delay dimension energy profile of $\mathbf{\hat{D}}$ with red color being the data symbols and green color representing the pilot.
After observing part a), where $P=1$ and $k_{\tau'} = 0$, we see the pilot at its expected position $p_0$. In part b), there is an energy leakage of data to the pilot (and vice versa) caused by a fractional delay. 
In part c), there are two channel paths with the maximum possible delay shift, yet both delays are integers, so no leakage is observed. 
Finally, part d) presents the worst case, where two channel paths with fractional delays cause severe energy leakage. 
Therefore, the delay guard size $M_{\rm g}$, as well as the extraction region size $M_{\rm ce}$ and its offset $m_{\rm ce}$, must be optimized for minimum interference, based on the expected channel behavior.

It is clear that a circularly pulse-shaped OTFS with an impulse pilot suffers notable interference due to interaction of the pulse shaping and channel. 
That can be moderated by the roll-off factor $\alpha$ (for the price of increased excess bandwidth) and the delay guard parameters $M_{\rm g}$, $M_{\rm ce}$ and $m_{\rm ce}$ (for the price of increased redundance). 
Therefore, we propose a design that prevents the delay leakage issues altogether, by adopting benefits of the UW-OFDM modulator \cite{b_Nonsys_OFDM}.

\section{Proposed UW-OTFS}

In this section, we present a solution to the delay energy leakage issue in CP-OTFS. As demonstrated in the previous section, the widely adopted impulse pilot channel estimation technique involves unfortunate trade-offs due to its operation after Rx pulse shaping and downsampling.

One possible remedy is to estimate the CSI before these steps, that is, in the oversampled time domain. However, at this stage, the pilot in CP-OTFS has already been dispersed by Tx pulse shaping, necessitating an alternative design.

Previously proposed UW-OFDM \cite{b_Huemer_UWODFM_1, b_Nonsys_OFDM} has a key advantage: its oversampled Tx waveform contains a zero guard interval of $M_{\rm gi}$ samples at the end of each OFDM symbol in the oversampled time domain. This feature allows for the inclusion of an arbitrary UW pilot within the guard interval, regardless of the pulse-shaping function. On the Rx side, the UW pilot can be readily sampled and used for CSI estimation.

Additionally, each delay block in UW-OFDM (a.k.a. an OFDM symbol) naturally exhibits smooth edges, which ensures minimal OOB emissions. Although UW-OFDM is not widely adopted in communication standards due to the higher receiver complexity compared to classical OFDM, its integration into the OTFS framework, as we proposed in our UW-OTFS, makes its advantages more prominent.

In this section, we start the UW-OTFS design description by definition of the input-output relation of the proposed UW-OTFS, assuming perfect CSI knowledge.

\subsection{Transmitter}

Similarly to CP-OTFS, the input $M \times N$ matrix, $\mathbf{D}$, is defined in the DD domain, only here, no pilot is embedded among the m-QAM data symbols.
The FT domain matrix is obtained using ISFT, $\mathbf{X} = \mathbf{F}_M \mathbf{D} \mathbf{F}_N^{\rm H}$,
which is then fed into the UW-OFDM modulator \cite{b_Nonsys_OFDM}. 
UW-OFDM now applies its precoding $\mathbf{G}$ to obtain
the $M_{\rm s} \times N$ FT domain precoded matrix $\mathbf{\check{X}} = \mathbf{G}\mathbf{X}$.
Here, $\mathbf{G}$ is the so-called "non-systematic" precoding matrix of size $M_{\rm s} \times M$ that is pre-calculated offline. 
$\mathbf{G}$ was first introduced in \cite{b_Nonsys_OFDM} as the case-2 matrix $\mathbf{\breve{G}}''$, which automatically adds $M_{\rm r}$ redundant subcarriers to the $M$ data subcarriers. As a consequence, an all-zero GI of length $M_{\rm gi}$ is created in the oversampled time domain - see waveform a) in Fig.~\ref{fig_UW_OTFS_waveforms}).
Note that $M_{\rm s}$ is the number of active subcarriers and $M_{\rm r}=M_{\rm s}-M$, where in our case $M_{\rm r}=M_{\rm gi}$. 
After precoding with $\mathbf{G}$, oversampling to size $M'$ by zero-padding is performed with a sparse matrix $\mathbf{B}$ of size $M' \times M_{\rm s}$, with the elements $B \big[\mathcal{M}_{\rm s}[p],p \big] = 1$ for $p \in \{0,1,\dots, M_{\rm s}-1 \}$ and where $\mathcal{M}_{\rm s}$ is the set of active subcarrier indices, defined as (for odd values of $M_{\rm s}$)
\begin{equation}\label{eq_M_s_set_WLAN_def}
\mathcal{M}_{\rm s} \coloneqq \Big\{0, 1,\dots, \frac{M_{\rm s}-1}{2}, M'-\frac{M_{\rm s}-1}{2}, \dots, M'-1 \Big\}.
\end{equation}
While the excess subcarriers in CP-OTFS are used for the RRC pulse shaping, in UW-OTFS they are generated by precoding with $\mathbf{G}$.
The $M' \times N$ UW-OTFS matrix in the oversampled delay-time domain is given by
\begin{equation}\label{eq_TX_s_mat_waveform_1}
    \mathbf{S}_{\rm d} = \alpha_{\rm d} \mathbf{F}_{M'}^* \mathbf{B} \mathbf{G}\mathbf{X} = \begin{pmatrix} \mathbf{\hat{S}}_{\rm d}^{\rm T} & \mathbf{0}_{M_{\rm gi}\times N}^{\rm T}
    \end{pmatrix}^{\rm T} ,
\end{equation}
where $\mathbf{0}_{M_{\rm gi}\times N}$ is the GI matrix of zeros, occurring naturally due to the precoding $\mathbf{G}$ and zero padding $ \mathbf{B}$.
The UW pilot (which we will define in the next section) is then added to the existing waveform producing the UW-OTFS transmit signal matrix $\mathbf{S}_{\rm uw} = \mathbf{S}_{\rm d} + \sigma_{\rm u}\mathbf{C}$, where $\mathbf{C}$ is a unit-energy UW pilot matrix, $\sigma_{\rm u}^2$ defines the mean UW pilot energy, and $\alpha_{\rm d}$ is the Tx power normalization coefficient for the data part, defined as
\begin{equation}\label{eq_alpha_def}
\alpha_{\rm d}^2 = M'(1 - \sigma_{\rm u}^2) /\tr \big\{\mathbf{G}^{\rm H} \mathbf{G} \big\}.
\end{equation}
The UW-OTFS data matrix in \eqref{eq_TX_s_mat_waveform_1} can also be expressed in terms of a signal $s_{\rm d}[k] = S_{\rm d} \big[(k)_{M'}, \floor{k/M'} \big]$ for $k \in [ 0, M'N-1 ]$ and zero otherwise, evaluated as
\begin{equation}\label{eq_TX_s_0_waveform_1}
    s_{\rm d}[k]  = \alpha_{\rm d} \sum_{n'=0}^{N-1} \sum_{m'=0}^{M_{\rm s}-1} \check{X}[m',n'] g_{\rm d}[k-n'M'] e^{j2\pi \mathcal{M}_{\rm s}[m'] \frac{k}{M'}},
\end{equation}
where the data part discrete window function is defined as $g_{\rm d}[k] = 1/\sqrt{M'}$ for all $k \in \{ 0, 1, \dots, M'-M_{\rm gi}-1 \}$ and zero otherwise.

\subsection{Equivalent Channel Matrix}

After propagating $s_{\rm d}[\cdot]$ through the LTV channel \eqref{eq_yk_channel_def}, the signal $r[\cdot]$ is received on the Rx input.
The main signal path of UW-OTFS then follows with DFT that produces an $M' \times N$ FT domain matrix $\mathbf{Y}$, defined as 
\begin{equation}\label{eq_R_0_RX_v1}
 Y[m,n] =  \sum_{k = -\infty}^{\infty} g_{0}[k-nM'] r[k] e^{-j2\pi \frac{mk}{ M'}}.
 \end{equation}
At this point, similarly to CP-OTFS, we may express each delay block of $M'$ samples separately. 
The $M' \times M$ ECM of the $n$-th delay block, $\mathbf{H}_n$, is defined as
\begin{equation}\notag
H_n[p,q] = \frac{\alpha_{\rm d}}{M'} \sum_{l=0}^{M_{\rm s}-1} G[l,q]   \sum_{i = 0}^{P-1} h[i] e^{-j2\pi \frac{p k_{\tau'}[i]}{M'}} 
\end{equation}
\begin{equation}\label{eq_UW_perf_ECM_def}
\times  e^{j2\pi n \xi[i]}  e^{j2\pi \frac{\xi[i] k_{\tau'}[i]}{ M'}}   \sum_{k = 0}^{M'-M_{\rm gi}-1}  e^{j2\pi \frac{ k}{ M'} \big( \mathcal{M}_{\rm s}[l] -p  + \xi[i] \big) }.
\end{equation}
The $n$-th delay block in the FT domain (i.e., the $n$-th column of $\mathbf{Y}$) is now evaluated as
\begin{equation}\label{eq_UW_R_vec_def}
\mathbf{y}_{n} =  \mathbf{H}_{n}\mathbf{x}_n + \mathbf{w}_n.
\end{equation}
where $\mathbf{w}_{n}$ is the AWGN vector and $\mathbf{x}_{n}$ is the $n$-th column of $\mathbf{X}$.

\subsection{Data Estimation}

In contrast to most of the existing OTFS literature, data estimation in UW-OTFS is performed in the FT domain, i.e., separately for each delay block $\mathbf{y}_{n}$ (e.g., similar to \cite{b_TimeDom_CE}), using the traditional LMMSE approach. 
Please note that even though the data signal estimation is done in the FT domain, the resulting FT matrix still needs to be transformed back to the DD domain with ISFT which allows to exploit the diversity gain of the doubly dispersive channel.

Before data estimation, it is worth pointing out that the spectral sidelobes of $\mathbf{y}_{n}$ fall very fast (see the black curve in the top part of Fig.~\ref{fig_EP_sym_energy_DD}). 
This may be used to reduce the complexity of the data estimation, so that instead of total $M'$ frequency bins of $\mathbf{y}_{n}$ we only use $M_{\rm b}$ strongest bins (where $M_{\rm b} \geq M_{\rm s}$). 
Their indices are defined as (for even values of $M_{\rm b}$)
\begin{equation}\label{eq_DE_size_reduction}
\mathcal{M}_{\rm b} \coloneqq \{0, 1, \dots M_{\rm b}/2-1, M'-M_{\rm b}/2, \dots, M'-1\},
\end{equation}
and (for odd values of $M_{\rm b}$)
\begin{equation}\label{eq_DE_size_reduction_odd}
\mathcal{M}_{\rm b} \coloneqq \Big\{0, 1,\dots, \frac{M_{\rm b}-1}{2}, M'-\frac{M_{\rm b}-1}{2}, \dots, M'-1 \Big\}.
\end{equation}
Therefore, for $m \in [ 0, M_{\rm b}-1 ]$, we have the reduced Rx vector given by $\breve{y}_{n}[m] = y_{n}[\mathcal{M}_{\rm b}[m]]$ and the reduced $n$th ECM, given by $\breve{H}_n[m,m'] = H_n[\mathcal{M}_{\rm b}[m],m']$, for $m' \in [ 0,M-1 ]$.
The $n$-th FT domain estimated data signal vector now yields 
\begin{equation}\label{eq_UW_X_est_vec_def}
\mathbf{\tilde{x}}_n = \mathbf{\Omega}_n\mathbf{\breve{y}}_{n},
\end{equation}
where $\mathbf{\Omega}_n$ is the $M \times M_{\rm b}$ LMMSE data estimator matrix ($M_{\rm b} > M$), defined as
\begin{equation}\label{eq_UW_Omega_def}
\mathbf{\Omega}_n = \big(\mathbf{\breve{H}}_n^{\rm H}\mathbf{\breve{H}}_n + \sigma_{w}^2\mathbf{I}_M \big)^{-1} \mathbf{\breve{H}}_n^{\rm H}.
\end{equation}
Finally, the $M \times N$ matrix of DD domain m-QAM symbol estimates for detection is given by stacking the FT domain estimates $\mathbf{\tilde{x}}_n$ into a matrix and performing SFT,
\begin{equation}\label{eq_UW_x_est_DD_mat_def}
\mathbf{\tilde{D}} = \mathbf{F}_M^{\rm H}\sum_{n=0}^{N-1} \mathbf{\tilde{x}}_n \mathbf{e}_n^{\rm T} \mathbf{F}_N.
\end{equation}
Here, we present the Tx and Rx structures of UW-OTFS while assuming perfect CSI knowledge on the Rx side. 
However, in the real application, the CSI needs to be estimated on the Rx side. The CSI estimation (CE) technique is proposed in the next section.

\begin{figure*}[t]
    \centering
    \includegraphics[width=\textwidth]{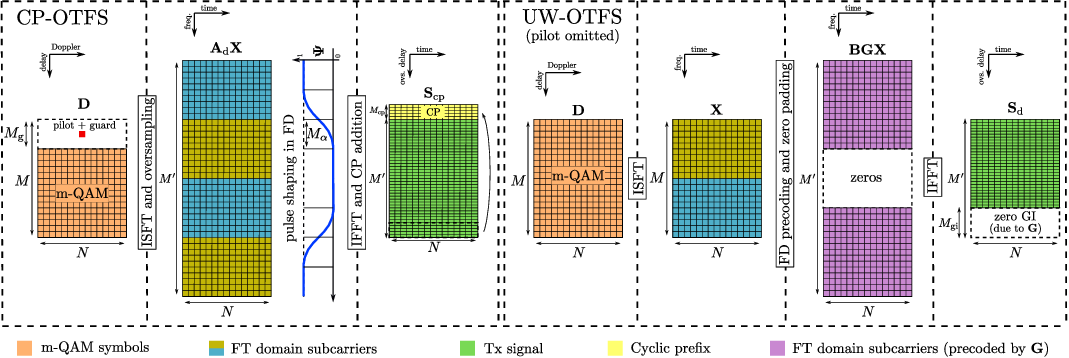}
    \caption{Transmitter structures of CP-OTFS and UW-OTFS (UW pilot omitted).}
\end{figure*}

\section{Proposed Pilot Design and CSI Estimation}

The CP-OTFS embedded pilot method described in Section IV estimates CSI in the DD domain, where a trade-off arises between data interference (i.e., CE quality) and bandwidth efficiency.

In contrast, UW-OTFS performs CE in the oversampled time domain, ensuring that the channel estimation process is not affected by data symbols. 
Here, we propose a novel CE method that is first illustrated using a sequence of Dirac impulses, which has been shown, via Proposition 2 in \cite{b_optimal_LTV_CE}, to be optimal for LTV channel estimation.

Although the Dirac sequence provides superior CE performance, its wideband spectral content makes it suboptimal for practical systems. We thus introduce a new UW pilot that offers lower OOB emissions at the cost of only slightly reduced CE quality.

\subsection{CSI Estimation in Oversampled Time Domain}

In the oversampled time domain, every delay block contains GI - the sequence of $M_{\rm gi}$ zero samples - see part a) of Fig.\ref{fig_UW_OTFS_waveforms}.
Since $M_{\rm gi}$ is an odd number, an impulse pilot can therefore to be placed in the center of every GI, forming a sequence of $N$ discrete Dirac impulses defined for all $k \in \mathbb{Z}$ as
\begin{equation}\label{eq_TX_s_u_ideal_waveform_1}
   s_{\rm u}[k]  = \sigma_{\rm u} \sqrt{M'}\sum_{n'=0}^{N-1}  \delta \big[k- n'M' - k_{\rm h}\big],
\end{equation}
where $k_{\rm h} = M'-M_{\rm h}$ and $M_{\rm h} = (M_{\rm gi}+1)/2$.
After propagating \eqref{eq_TX_s_u_ideal_waveform_1} through the LTV channel \eqref{eq_yk_channel_def}, we obtain (without AWGN)
\begin{equation}\label{eq_yu_dirac_channel_def}
y_{\rm u}[k] = \sigma_{\rm u} \sqrt{M'} \sum_{i = 0}^{P -1}  h[i] \sum_{n'=0}^{N-1} \delta \big[k-k_{\tau'}[i]-n'M'-k_{\rm h} \big]   e^{j2\pi \frac{\xi[ i] k}{M'} }.
\end{equation}
Our proposed CE method now follows by extracting the last $M_{\rm h}$ samples of each received delay block and stacking them into the $M_{\rm h} \times N$ CE matrix, defined as 
\begin{equation}\label{eq_UW_y_ce_mat_def_0}
Y_{\rm ce}[m, n] =  y_{\rm u}[m + M'n +k_{\rm h}],
\end{equation}
which, after including \eqref{eq_yu_dirac_channel_def}, is expanded into
\begin{equation}\label{eq_UW_y_ce_mat_def}
Y_{\rm ce}[m, n] = \sigma_{\rm u} \sqrt{M'}\sum_{i = 0}^{P -1}  h[i]  \delta \big[m - k_{\tau'}[i]\big]   e^{j2\pi \xi[ i] \frac{m + M'n +k_{\rm h}}{M'} }.
\end{equation}
As long as the maximum oversampled channel memory $L'$ satisfies the following condition
\begin{equation}\label{eq_UW_Channel_Memory}
M_{\rm gi} \geq 2L' -1,
\end{equation}
CSI can be estimated using the last $M_{\rm h}$ samples of each delay block, without interference from data.

Unlike in \cite{b_TFdomain_OTFS_CE}, our approach does not require knowledge of the number of channel taps, P, at the receiver. 
Given that the CSI is distributed among three separate parameters of unknown length $P$, i.e., $h[i]$, $k_{\tau'}[i]$ and $\xi[i]$, their direct estimation from $\mathbf{Y}_{\rm ce}$ is challenging.
Therefore, we present a modified GCE-BEM technique, based on \cite{b_Farhang_BEM_ICC, b_BEM_Gonzales_WCL2024, b_BEM_Turbo_TCOMM2024, b_BEM_TCOMM2022}, to unify the CSI into a single matrix with known dimensions that can easily be estimated from $\mathbf{Y}_{\rm ce}$ using the LMMSE approach.

\subsection{Proposed CSI Estimation - Dirac Impulses}

According to Proposition 4 in \cite{b_optimal_LTV_CE}, the optimal number of training blocks should match the Doppler dimension size of the BEM coefficient matrix, i.e., $N$. 
Therefore, we proceed by quantizing the continuous Doppler domain $\nu \in [ -\nu_{\rm max}, \nu_{\rm max})$ into $N$ points, equally spaced by the nominal Doppler resolution $\Delta \nu = \Delta f/N$. 
In other words, $\nu \to k_\nu \Delta \nu$, where $k_\nu \in \mathcal{V}$ and where 
\begin{equation}\label{eq_k_nu_UW_def_alt}
 \mathcal{V} \coloneqq \big\{ 0, 1, \dots, N-1 \big\} - (N-1)/2.
\end{equation}
The CSI is now contained within an $M_{\rm h} \times N$ matrix of complex channel gains, $\mathbf{H}_{\rm ce}$, a.k.a. the BEM coefficient matrix.
We now introduce the GCE-BEM modification of the signal propagated through the LTV channel in \eqref{eq_yk_channel_def} as
\begin{equation}\label{eq_yu_k_sampled_DD_def}
r[k]  = \sum_{k_{\tau'} = 0}^{M_{\rm h} -1} s[k-k_{\tau'}]\sum_{k_{\nu} =0}^{N-1} H_{\rm ce}[k_{\tau'},k_{\nu} ]  e^{j2\pi  
 \frac{\mathcal{V}[k_{\nu}] k}{n_\nu N M_{\rm x}'} } + w[k],
\end{equation}
where $n_\nu$ is the GCE-BEM rate factor and where $M_{\rm x}' = M'$.
The CE matrix \eqref{eq_UW_y_ce_mat_def_0} is now evaluated using the Dirac impulse UW sequence \eqref{eq_TX_s_u_ideal_waveform_1}, as (AWGN omitted for clarity)
\begin{equation}\label{eq_UW_y_ce_mat_sampled_def}
Y_{\rm ce}[m, n] = \sigma_{ \rm u} \sqrt{M'} \sum_{k_{\nu} =0}^{N-1} H_{\rm ce}[m,k_{\nu} ]    e^{j2\pi \frac{\mathcal{V}[k_{\nu}]}{n_\nu} \frac{m + M'n +k_{\rm h}}{ M'N} }.
\end{equation}
Unlike in \eqref{eq_UW_y_ce_mat_def}, the CSI in \eqref{eq_UW_y_ce_mat_sampled_def} has the same dimensions as $\mathbf{Y}_{\rm ce}$, thus, forming a linear system of equations from which $\mathbf{H}_{\rm ce}$ may be estimated based on the LMMSE criterion.

In the GCE-BEM literature, the rate factor $n_\nu$ is exclusively set as a positive integer \cite{b_Farhang_BEM_ICC, b_BEM_Gonzales_WCL2024, b_BEM_Turbo_TCOMM2024, b_BEM_TCOMM2022}.  
However, provided that the maximum expected Doppler spread $\nu_{\rm max}$ is known to the receiver, $n_\nu$ can be used to scale the GCE-BEM Doppler grid $\mathcal{V}$ arbitrarily and maximize the CE performance of UW-OTFS.
As it will be numerically shown in Section X, for a given $\nu_{\rm max}$ and $\Delta f$, the optimal GCE-BEM rate approaches 
\begin{equation}\label{eq_optimal_BEM_rate}
n_{\nu_0} \cong \Delta f/(2 \nu_{\rm max}).
\end{equation}

\subsection{Proposed UW Pilot}

\begin{figure}[t!] \centering
\includegraphics[scale=0.55]{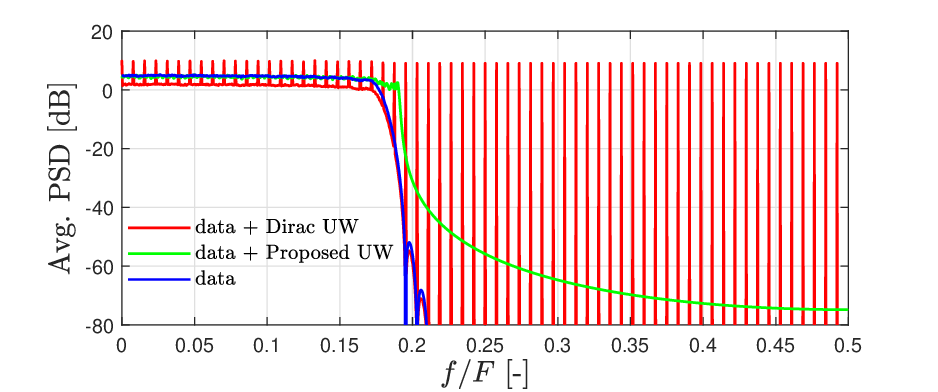}
\caption{PSD comparison of the UW-OTFS data transmission $s_{\rm d}(.)$ in \eqref{eq_TX_s_0_waveform_1} with the Dirac UW pilot \eqref{eq_TX_s_u_ideal_waveform_1}, with the proposed UW pilot \eqref{eq_TX_s_u_waveform_1} and without the pilot. }
\label{fig_PSD_UW}
\end{figure}

Although being optimal in the CE sense \cite{b_optimal_LTV_CE}, the sequence of Dirac impulses has unlimited bandwidth. 
This severely degrades the otherwise low spectral OOB emissions of the UW-OFDM modulator \cite{b_Nonsys_OFDM} - see Fig.~\ref{fig_PSD_UW}. 
Therefore, it is essential to preserve the naturally low OOB emissions of the UW-OFDM at the cost of slightly reducing the CE quality.  

The proposed UW pilot was inspired by the Dirac impulse pilot in its flat PSD shape. The novelty lies in populating only $M_{\rm s}$ of the total $M'$ subcarriers, defined by the subcarrier indices set in \eqref{eq_M_s_set_WLAN_def}, which limits the OOB emissions of the resulting UW pilot.
If viewed in the oversampled time domain, the UW pilot forms a vector $\mathbf{c}_0$ of size $M'N \times 1$, defined as
\begin{equation}\label{eq_UW_c_vec_def}
c_0[p] =  \frac{1}{\sqrt{M'- M_{\rm s}}} \sum_{k \in \mathcal{M}_{\rm s}} e^{j 2 \pi k \frac{p + M_{\rm h}}{M'}}.
\end{equation}
It resembles the Dirichlet kernel function with a period $M'$ and a single dominant peak shifted by $M_{\rm h}$ samples to the center of GI to minimize contamination by data samples and vice versa - see waveform b) in Fig.~\ref{fig_UW_OTFS_waveforms} ($97.2 \%$ of total UW energy lies within GI, given that $M'=128$ and $M_{\rm gi}=17$). 
However, since only the last $M_{\rm h}$ samples of GI are extracted for CE purposes in \eqref{eq_UW_y_ce_mat_def_0}, and because the novel UW pilot energy is spread over the whole delay block, it suffers performance loss compared to the Dirac impulse pilot.  

Due to periodicity in the oversampled time domain, the novel UW pilot contains PSD spikes that are more than $10$ dB above the mean PSD level. If left untreated, such PSD artifacts would limit the average Tx power of UW-OTFS due to spectral mask violations.
This problem was solved (without any performance penalty) by multiplying $\mathbf{c}_0$ by a linear chirp sequence with a period of $M'N$ samples.
The linear chirp, having a constant magnitude, only affects the phase of the pilot; therefore, its period extends from $M'$ to $M'N$ samples. 
We thus define the resulting UW pilot vector, $\mathbf{c}$, as
\begin{equation}\label{eq_UW_c_vec_def2}
c[p] = \frac{e^{-j\pi \frac{p}{M'}\big(\frac{p}{M'N} -1 \big)}}{\sqrt{M'- M_{\rm s}}}  \sum_{k \in \mathcal{M}_{\rm s}} e^{j 2 \pi k\frac{p + M_{\rm h}}{M'}}.
\end{equation}

\subsection{Proposed CSI Estimation - Arbitrary UW Pilot}

Formula \eqref{eq_TX_s_u_ideal_waveform_1} can be extended to accommodate an arbitrary UW pilot vector of size $M'N \times 1$, including our vector $\mathbf{c}$ defined in \eqref{eq_UW_c_vec_def2}, and it yields  
\begin{equation}\label{eq_TX_s_u_waveform_1}
   s_{\rm u}[k]  = \sigma_u \sqrt{M'} \sum_{n'=0}^{N-1} c\big[ (k)_{M'} + n'M' \big] g_{0}[k-n'M'],
\end{equation}
where $g_{0}[\cdot]$ is a window function defined in \eqref{eq_g0_window_def}.
Once $s_{\rm u}[\cdot]$ passes through the GCE-BEM LTV channel in \eqref{eq_yu_k_sampled_DD_def}, we obtain $y_{\rm u}[\cdot]$ (AWGN omitted for clarity) and the CE matrix in \eqref{eq_UW_y_ce_mat_def_0} is evaluated as
\begin{equation}\label{eq_y_ce_mat_sampl_def}
Y_{\rm  ce}[m, n] =   \sum_{k_{\tau'} = 0}^{M_h - 1} \sum_{k_{\nu} = 0}^{N-1} \bar{A}_{\rm ce}[m, n, k_{\tau'}, k_{\nu}]H_{\rm ce}[k_{\tau'},k_{\nu}],
\end{equation}
where the $M_{\rm h} \times N \times M_{\rm h} \times N$ tensor $ \mathbf{\bar{A}}_{\rm ce}$ is defined as
\begin{equation}\notag
\bar{A}_{\rm ce}[m, n, k_{\tau'}, k_{\nu}] =  \sigma_{\rm u}  e^{j2\pi \frac{\mathcal{V}[k_{\nu}] (m+ M'n+k_{\rm h})}{n_\nu N M'}}
\end{equation}
\begin{equation}\label{eq_UW_A_ce_def_simpl}
\times c \big[ (m - k_{\tau'} - M_{\rm h})_{M'} + nM' \big].
\end{equation}
The relation in \eqref{eq_y_ce_mat_sampl_def} can be reshaped into a linear system (after adding the AWGN matrix $\mathbf{W}_{\rm ce}$)
\begin{equation}\label{eq_r_ce_vec_def}
\mathbf{y}_{\rm ce} =   \mathbf{A}_{\rm ce} \mathbf{h}_{\rm ce} + \mathbf{w}_{ce},
\end{equation}
where $\mathbf{y}_{\rm ce} \equiv \vect\{\mathbf{Y}_{\rm ce} + \mathbf{W}_{\rm ce}\}$, $\mathbf{w}_{\rm ce} \equiv \vect\{ \mathbf{W}_{\rm ce}\}$,  $\mathbf{h}_{\rm ce} \equiv \vect\{\mathbf{H}_{\rm ce}\}$ and where the $M_{\rm h}N \times M_{\rm h}N$ matrix $\mathbf{A}_{\rm ce}$ is defined as
\begin{equation}\label{eq_UW_A_ce_mat_def}
A_{\rm ce}[p,q] \equiv \bar{A}_{\rm ce}\Big[(p)_{M_{\rm h}}, \floor[\big]{p/M_{\rm h}}, (q)_{M_{\rm h}}, \floor[\big]{q/M_{\rm h}} \Big],
\end{equation}
for $p,q \in \{0,1,\dots, M_{\rm h}N-1\}$.
The GCE-BEM coefficient vector estimate $\mathbf{\tilde{h}}_{\rm ce}$ can be obtained with the LMMSE estimator
\begin{equation}\label{eq_CE_LMMSE}
\mathbf{\tilde{h}}_{\rm ce} =  \mathbf{\Omega}_{\rm ce} \mathbf{y}_{\rm ce},
\end{equation}
where $\mathbf{\Omega}_{\rm ce} = \mathbf{A}_{\rm ce}^{\rm H}\big(\mathbf{A}_{\rm ce} \mathbf{A}_{\rm ce}^{\rm H} + \gamma \mathbf{I}_{M_{\rm h}N} \big)^{-1}$, and where $\gamma = \sigma_{\rm w}^2 M_{\rm h}N$.
The estimated GCE-BEM coefficient matrix of size $M_{\rm h} \times N$ is then given by $\mathbf{\widetilde{H}}_{\rm ce} = \vect^{-1}\{ \mathbf{\tilde{h}}_{\rm ce}\}$. 
Since $\mathbf{A}_{\rm ce}$ contains no CSI, it can be calculated offline, and the noise variance $\sigma_{\rm w}^2$ remains the only variable in the matrix inversion. 
Therefore, $\mathbf{\Omega}_{\rm ce}$ can be simplified (using eigendecomposition) into 
\begin{equation}\label{eq_UW_Omega_CE_def}
\mathbf{\Omega}_{\rm ce} = \mathbf{A}_{\rm ce}^{\rm H}\mathbf{Q} \big(\mathbf{\Lambda}_{\rm ce} +  \gamma \mathbf{I}_{M_{\rm h}N} \big)^{-1} \mathbf{Q}^{\rm H},
\end{equation}
 where $\mathbf{\Lambda}_{\rm ce}$ is a real-valued diagonal matrix containing the eigenvalues of $\mathbf{A}_{\rm ce} \mathbf{A}_{\rm ce}^{\rm H}$ and where $\mathbf{Q}$ is the matrix of eigenvectors. 
 The complex-valued matrix inversion of size $M_{\rm h}N \times M_{\rm h}N$ is therefore reduced into a real-valued $M_{\rm h}N \times 1$ vector inversion, where $\mathbf{A}_{\rm ce}^{\rm H}\mathbf{Q} $, $\mathbf{\Lambda}_{\rm ce}$ and $\mathbf{Q}^{\rm H}$ can all be calculated offline.

\subsection{ECM and Data Estimation}

To calculate the ECM using estimated CSI, we must follow the same approach as presented in \eqref{eq_UW_perf_ECM_def}, however, the LTV channel input-output relation now employs the GCE-BEM formula \eqref{eq_yu_k_sampled_DD_def} instead of \eqref{eq_yk_channel_def}. 
The $M' \times M$ ECM of the $n$-th delay block, $\mathbf{\widetilde{H}}_n$, is then given by
\begin{equation}\label{eq_UW_est_ECM_def}
\widetilde{H}_n[p, q] = \sum_{k_{\nu} =0}^{N-1}  
 \bar{G} [p,q,k_\nu] N[n, k_\nu]
 M[p, k_{\nu}],
\end{equation}
where
\begin{equation}\label{eq_UW_Complexity_M_mat_def}
M[p, k_{\nu}] = 
   \sum_{k_{\tau'} = 0}^{M_{\rm h}-1} \bar{M}[p, k_{\tau'}, k_{\nu}] \widetilde{H}_{\rm ce}[k_{\tau'},k_{\nu}],
\end{equation}
and where the offline-computed tensors are defined by
\begin{equation}\label{eq_UW_Complexity_M_tensor_def}
\bar{M}[p, k_{\tau'}, k_{\nu}] = 
 e^{-j2\pi \frac{p  k_{\tau'} }{ M'}}   e^{j2\pi \frac{\mathcal{V}[k_{\nu}] k_{\tau'} }{n_\nu N M'}},
\end{equation}
\begin{equation}\label{eq_UW_Complexity_G_tensor_def}
\bar{G} [p,q,k_\nu] = \frac{\alpha}{M'}   \sum_{l=0}^{M_s-1} G[l,q]  \sum_{k = 0}^{M'-M_{gi}-1} e^{j2\pi k\frac{\mathcal{M}_s[l] - p + \frac{\mathcal{V}[k_{\nu}]}{n_\nu N}}{M'} },
\end{equation}
and
\begin{equation}\label{eq_UW_Complexity_N_mat_def}
N[n, k_\nu] = e^{j2\pi \frac{\mathcal{V}[k_{\nu}] n}{n_\nu N}}.
\end{equation}
Similarly to \eqref{eq_UW_X_est_vec_def}, the size of the ECM in \eqref{eq_UW_est_ECM_def} can be reduced from $M' \times M$ to $M_{\rm b} \times M$, according to $\breve{H}_n[m,m'] = \widetilde{H}_n[\mathcal{M}_b[m],m']$, for $m \in [ 0,M_b-1 ]$ and $m' \in [ 0,M-1 ]$.
Lastly, the LMMSE matrix in \eqref{eq_UW_Omega_def} is used for data estimation and the SFT in \eqref{eq_UW_x_est_DD_mat_def} then yields the m-QAM symbol estimates in the DD domain.
Algorithm~\ref{ALG_UW} summarizes the Rx processing of UW-OTFS.

\begin{algorithm} 
    \caption{Receiver processing of UW-OTFS} 
    \label{ALG_UW}
    \begin{flushleft}
        \textbf{INPUT:} \hangindent=1.6em \hangafter=1
        \parbox[t]{0.9\linewidth}{UW-OTFS Rx signal $r[\cdot]$ in \eqref{eq_yk_channel_def}, $\mathbf{A}_{\rm ce}$ in \eqref{eq_UW_A_ce_mat_def}, $\mathbf{A}_{\rm ce}^{\rm H}\mathbf{Q}$, $\mathbf{\Lambda}_{\rm ce}$, and $\mathbf{Q}^{\rm H}$ from \eqref{eq_UW_Omega_CE_def}, tensors \eqref{eq_UW_Complexity_M_tensor_def}, \eqref{eq_UW_Complexity_G_tensor_def}, and \eqref{eq_UW_Complexity_N_mat_def}.} \\ \vspace{0.3em} 
        \textbf{OUTPUT:} \parbox[t]{0.9\linewidth}{Matrix of DD domain m-QAM symbol estimates, $\mathbf{\tilde{D}}$.}
    \end{flushleft}
    \begin{algorithmic}[1]
        \State Extract the CE regions from each delay block of $r[\cdot]$ to $\mathbf{Y}_{\rm ce}$ with \eqref{eq_UW_y_ce_mat_def_0}.
        \State Calculate $\mathbf{\Omega}_{\rm ce}$ with \eqref{eq_UW_Omega_CE_def}.
        \State Calculate the GCE-BEM vector estimate $\mathbf{\tilde{h}}_{\rm ce}$ with \eqref{eq_CE_LMMSE}. Reshape it to a matrix using $\mathbf{\widetilde{H}}_{\rm ce} = \vect^{-1}\{ \mathbf{\tilde{h}}_{\rm ce}\}$.
        \State Use Wigner transform in \eqref{eq_R_0_RX_v1} to obtain $\mathbf{Y}$ from the input signal $r[\cdot]$. 
        \State Reduce the size with $\breve{y}_{n}[m] = Y[\mathcal{M}_{\rm b}[m],n]$ for $m \in [ 0, M_{\rm b}-1 ]$.
        \State Calculate $\mathbf{M}$ using \eqref{eq_UW_Complexity_M_mat_def}. 
        \ForAll {$n$ = 0 to $N-1$}
            \State \parbox[t]{0.9\linewidth}{Calculate $\breve{H}_n[m,m'] = \widetilde{H}_n[\mathcal{M}_b[m],m']$ using \eqref{eq_UW_est_ECM_def}, for $m \in [ 0, M_{\rm b}-1 ]$, $m' \in [ 0, M-1 ]$.} \vspace{0.2em} 
            \State Use \eqref{eq_UW_Omega_def} to obtain $\mathbf{\Omega}_n$.
            \State Calculate the data estimate vector $\mathbf{\tilde{x}}_n$ with \eqref{eq_UW_X_est_vec_def}.
        \EndFor
        \State Perform SFT with \eqref{eq_UW_x_est_DD_mat_def} to obtain $\mathbf{\tilde{D}}$.
    \end{algorithmic}
\end{algorithm}

\section{CP-OTFS with Estimated CSI}

To benchmark our UW-OTFS concept, we present a complete Rx processing for CP-OTFS as introduced in Sections III and IV-A for the case of estimated CSI knowledge.

\subsection{Equivalent Channel Matrix and Data Estimation}

To ensure that the comparison between CP-OTFS and UW-OTFS is carried out in a fair manner, we propose the same modified GCE-BEM for CP-OTFS. 

The $M' \times N$ matrix of the  received signal in the FT domain, $\mathbf{Y}$, is defined in
\eqref{eq_mCP_Y_mat_eval_4}. 
We now replace the $n$-th ECM, $\mathbf{H}_n$, by the estimated CSI ECM, $\mathbf{\widetilde{H}}_n$. 
Similarly to its UW-OTFS counterpart \eqref{eq_UW_est_ECM_def}, $\mathbf{\widetilde{H}}_n$ in CP-OTFS uses the modified GCE-BEM LTV input-output channel relation \eqref{eq_yu_k_sampled_DD_def}, which yields
\begin{equation}\label{eq_ECT_BEM_def}
\widetilde{H}_n[m,m'] =  \sum_{k_{\nu} =0}^{N-1} N[n, k_\nu]   \sum_{k=0}^{Q-1} 
\bar{G}[m, m', k_\nu, k] \bar{M}[m',k_{\nu},k],  
\end{equation}
where
\begin{equation}\label{eq_CP_M_tensor_def}
\bar{M}[m',k_{\nu},k] =  \sum_{k_{\tau'} = 0}^{M_h -1}  \widetilde{H}_{\rm ce}[k_{\tau'},k_{\nu} ]\bar{M}_0[m', k_{\tau'},k] ,  
\end{equation}
and where $M_{\rm h} = M_{\rm cp} + 1$. 
The offline-calculated tensors are defined as
\begin{equation}\label{eq_Complexity_Mk_mat_def}
\bar{M}_0[m', k_{\tau'},k] =  e^{-j2\pi \frac{(m'+kM) k_{\tau'} }{ QM} },
\end{equation}
\begin{equation}\label{eq_Complexity_Gk_mat_def}
\bar{G}[m, m', k_\nu, k] =  \psi[m'+kM]    
\chi  \Big(m'+kM-m + \frac{\mathcal{V}[k_{\nu}] M' }{n_\nu N M_{\rm x}'}, M' \Big),
\end{equation}
where $N[n, k_\nu]$ is defined in \eqref{eq_UW_Complexity_N_mat_def}. 
Here, $M_{\rm x}' = M' + M_{\rm cp}$ and $\chi (\cdot, \cdot)$ is defined in \eqref{eq_mCP_Dirichlet_kern_def}.
Similarly to UW-OTFS, the size of the ECM in \eqref{eq_ECT_BEM_def} is reduced according to $\breve{H}_n[m,m'] = \widetilde{H}_n[\mathcal{M}_b[m],m']$. 
The LMMSE data estimator matrix is defined as \eqref{eq_UW_Omega_def} and the CP-OTFS Rx processing is finished by \eqref{eq_UW_x_est_DD_mat_def}, yielding the DD domain matrix of m-QAM symbol estimates, $\mathbf{\tilde{D}}$. 
Next, we show how to obtain the estimated GCE-BEM coefficient matrix, $\mathbf{\widetilde{H}}_{\rm ce}$.

\subsection{CSI Estimation}

As briefly introduced in Section IV, the CSI estimation in CP-OTFS begins with three steps: DFT of size $M'$, pulse shaping and aliasing back to size $M \times N$.
The resulting FT domain matrix yields $\mathbf{\hat{Y}} = \mathbf{A}_{\rm d}^{\rm T} \boldsymbol{\Psi} \mathbf{Y}$, and after SFT in \eqref{eq_mCP_y_DD_mat_eval}, we obtain the DD domain matrix $\mathbf{\hat{D}}$.
From $\mathbf{\hat{D}}$, we extract the region of size $M_{\rm ce} \times N$, defined as $Y_{\rm ce}[m,n] = \hat{D}[m+m_{\rm ce},n]$, where $M_{\rm ce}$, $M_{\rm g}$, and $m_{\rm ce}$ are chosen based on a numerical analysis and the knowledge from Section IV. 
The relation between the BEM coefficient matrix, $\mathbf{H}_{\rm ce}$, and the DD domain matrix, $\mathbf{Y}_{\rm ce}$, is given by (AWGN neglected)
\begin{equation}\label{eq_CP_BEM_CE_def}
Y_{\rm ce}[m,n] =  \sum_{k_{\tau'} = 0}^{M_h -1}  \sum_{k_{\nu} =0}^{N-1} H_{\rm ce}[k_{\tau'},k_{\nu} ] \bar{A}_{\rm ce} [m+m_{\rm ce},n,k_{\tau'} ,k_{\nu}],   
\end{equation}
where the pre-calculated tensor $\mathbf{\bar{A}}_{\rm ce}$ is given by
\begin{equation}\notag
\bar{A}_{\rm ce}[m,n,k_{\tau'} ,k_{\nu}] =  \frac{x_0\sqrt{Q}}{M } \chi \Big(q_0 - n + \frac{\mathcal{V}[k_{\nu}]}{n_\nu}, N \Big)  \sum_{l=0}^{M-1}e^{j2 \pi \frac{ml}{M}} 
\end{equation}
\begin{equation}\notag
 \sum_{m'=0}^{M-1} e^{-j2\pi \frac{m'(p_0+k_{\tau'}/Q)}{M}} \sum_{k=0}^{Q-1} \psi[l + kM] \sum_{k'=0}^{Q-1} \psi[m'+k'M]
\end{equation}
\begin{equation}\label{eq_mCP_A_CE_PS_Q_eval}
\times   e^{-j 2 \pi \frac{k_{\tilde{\tau}} k'}{Q} } 
\chi  \Big(m'-l + (k'-k)M + \frac{\mathcal{V}[k_{\nu}] M' }{n_\nu N M_{\rm x}'}, M' \Big).
\end{equation}
Note that $\mathbf{\bar{A}_{\rm ce}}$ in \eqref{eq_mCP_A_CE_PS_Q_eval} assumes that only the pilot is present in the DD domain matrix $\mathbf{D}$ in \eqref{eq_CP_x_DD_mat_def}. 
Similarly to \eqref{eq_UW_A_ce_mat_def}, we now reshape \eqref{eq_mCP_A_CE_PS_Q_eval}, only this time to a rectangular matrix of size $M_{\rm ce}N \times M_{\rm h} N$ with elements
\begin{equation}\label{eq_CP_A_ce_mat_def}
A_{\rm ce}[m,k] =  \bar{A}_{\rm ce}\big[ (m)_{M_{\rm ce}}, \floor{m/M_{\rm ce}}, (k)_{M_{\rm h}} ,\floor{k/M_{\rm h}} \big],
\end{equation}
where $m \in [ 0, M_{\rm ce}N-1 ]$ and $k \in [ 0, M_{\rm h} N-1 ]$.
The GCE-BEM coefficient matrix is obtained by calculating the LMMSE estimator matrix \eqref{eq_UW_Omega_CE_def} with the dimension of the diagonal matrix inversion being $M_{\rm ce}N \times M_{\rm ce}N$ and $\gamma = \sigma_w^2 M_{\rm ce}N$, followed by \eqref{eq_CE_LMMSE}, where $\mathbf{\widetilde{H}}_{\rm ce} = \vect^{-1}\{ \mathbf{\tilde{h}}_{\rm ce}\}$.
Algorithm~\ref{ALG_CP} summarizes the Rx processing of CP-OTFS.

\begin{algorithm} 
    \caption{Receiver processing of CP-OTFS} 
    \label{ALG_CP} 
    \begin{flushleft}
        \textbf{INPUT:} \hangindent=1.6em \hangafter=1
        \parbox[t]{0.9\linewidth}{CP-OTFS Rx signal $r[\cdot]$ in \eqref{eq_yk_channel_def}, $\mathbf{A}_{\rm ce}$ in \eqref{eq_CP_A_ce_mat_def}, $\mathbf{A}_{\rm ce}^{\rm H}\mathbf{Q}$, $\mathbf{\Lambda}_{\rm ce}$, and $\mathbf{Q}^{\rm H}$ from \eqref{eq_UW_Omega_CE_def}, tensors \eqref{eq_Complexity_Mk_mat_def}, \eqref{eq_Complexity_Gk_mat_def}, and \eqref{eq_UW_Complexity_N_mat_def}.} \\ \vspace{0.3em} 
        \textbf{OUTPUT:} \parbox[t]{0.9\linewidth}{Matrix of DD domain m-QAM symbol estimates, $\mathbf{\tilde{D}}$.}
    \end{flushleft}
    \begin{algorithmic}[1]
        \State Use Wigner transform in \eqref{eq_mCP_Y_mat_def} to obtain the FT domain matrix $\mathbf{Y}$.
        \State Apply Rx pulse shaping and aliasing to obtain $\mathbf{\hat{Y}} = \mathbf{A}_{\rm d}^{\rm T} \boldsymbol{\Psi} \mathbf{Y}$.
        \State Use SFT in \eqref{eq_mCP_y_DD_mat_eval} to obtain $\mathbf{\hat{D}}$.
        \State Extract the CE region using $Y_{\rm ce}[m,n] = \hat{D}[m+m_{\rm ce},n]$ for $m \in [ 0, M_{\rm ce} -1]$ and $n \in [ 0,N-1 ]$.
        \State Calculate $\mathbf{\Omega}_{\rm ce}$ with \eqref{eq_UW_Omega_CE_def}.
        \State Calculate the GCE-BEM vector estimate $\mathbf{\tilde{h}}_{\rm ce}$ with \eqref{eq_CE_LMMSE}. Reshape it to a matrix using $\mathbf{\widetilde{H}}_{\rm ce} = \vect^{-1}\{ \mathbf{\tilde{h}}_{\rm ce}\}$.
        \State Perform data estimation in the FT domain. First, calculate the tensor $\mathbf{\bar{M}}$, using \eqref{eq_CP_M_tensor_def}. 
        \State Perform the following for each delay block:
        \ForAll {$n$ = 0 to $N-1$}
        \State \parbox[t]{0.9\linewidth}{Calculate $\breve{H}_n[m,m'] = \widetilde{H}_n[\mathcal{M}_b[m],m']$ using \eqref{eq_ECT_BEM_def} for $m \in [ 0, M_{\rm b}-1 ]$, $m' \in [ 0, M-1 ]$.} \vspace{0.2em} 
        \State Use \eqref{eq_UW_Omega_def} to obtain $\mathbf{\Omega}_n$.
        \State Calculate the data estimate vector $\mathbf{\tilde{x}}_n$ with \eqref{eq_UW_X_est_vec_def}.
        \EndFor
        \State Perform SFT with \eqref{eq_UW_x_est_DD_mat_def} to obtain $\mathbf{\tilde{D}}$.
    \end{algorithmic}
\end{algorithm}

\section{Computational Complexity and Memory Footprint Analysis}

In this section, we derive the computational complexity and memory footprint expressions for CP-OTFS and for the proposed UW-OTFS. 
Similarly to \cite{b_Huemer_UWODFM_1}, the computational complexity analysis is done by counting the complex multipliers (CM), where complex divisions are assumed to be identical to the CM and complex additions are neglected. 
Offline computations, such as those in equation \eqref{eq_mCP_A_CE_PS_Q_eval}, are excluded from the complexity analysis.

\subsection{ISFT and SFT}

In both CP-OTFS and UW-OTFS, ISFT and SFT are utilized to transform the $M \times N$ matrix from the DD domain to the FT domain and vice versa. 
Assuming that both $M$ and $N$ are powers of $2$, ISFT and SFT can be done using the radix-2 fast Fourier transform (FFT) algorithm and the required CM count is $C_{\rm isft} = C_{\rm sft} = N M/2 \log_2(MN)$.

\subsection{CP-OTFS}

After ISFT, Tx processing in CP-OTFS continues by cyclically extending the FT domain matrix to size $M' \times N$ (where $M' = QM$) and by multiplication with the $M' \times M'$ real-valued diagonal matrix $\mathbf{\Psi}$, which requires $M'N/2$ CM (in reality it is less than $M'N/2$ CM, since only the s-curves of the pulse shaping filter require multiplications). 
Next, the IFFT of size $M'$ is performed ($N$ times), requiring $C_{\rm ifft} =  C_{\rm fft} = M'/2 \log_2(M')$ CM per delay block.

\renewcommand{\arraystretch}{1.4}
\begin{table}[t]
\caption{Computational Complexity Expressions}
\label{table:1}
\centering
\begin{tabular} {|c|c|c|} 
\hline\hline
 {} &  {CP-OTFS} &  {UW-OTFS}\\ \hline\hline
Tx side & $C_{\rm isft} + M Q N/2 + NC_{\rm ifft}$ & $C_{\rm isft} + M_s M N + NC_{\rm ifft}$ \\ \hline
$\mathbf{\Omega}_{\rm ce}$ & \multicolumn{2}{c|}{$ M_{\rm h}M_{\rm ce}^2 N^3 + 1/2 M_{\rm ce}^2 N^2 + 1/4 M_{\rm ce} N$} \\ \hline
$\mathbf{\tilde{H}}_{n}$ (all $N$) & $MN((M_{\rm h} + M_{\rm b})Q + M_{\rm b}N)$    & $M_{\rm b}N(M_{\rm h} + 2MN)$ \\ \hline
$\mathbf{\Omega}_{n}$ (all $N$) & \multicolumn{2}{c|}{$N(C_{\rm Herm} + C_{\rm inv})$}  \\ \hline
    \hline
\end{tabular}
\end{table}

On the Rx side, FFT of size $M'$ is performed ($N$ times), after which the signal path is split into the channel and the data estimation part.
For channel estimation, pulse shaping is performed with $\mathbf{\Psi}$, followed by aliasing and SFT, which together require $M'N/2 + N M/2 \log_2(MN)$ CM. 
In the DD domain, a section of size $M_{\rm ce} \times N$ is extracted from $\mathbf{\hat{D}}$, defined as $\mathbf{Y}_{\rm ce}$. 
The estimate of the BEM coefficient matrix, $\mathbf{\widetilde{H}}_{\rm ce}$, is then obtained from \eqref{eq_CE_LMMSE}, which requires $M_{\rm h}^2 N^2$ CM, and the computation of $\mathbf{\Omega}_{\rm ce}$ with the eigendecomposition method in \eqref{eq_UW_Omega_CE_def} requires $ M_{\rm h}M_{\rm ce}^2 N^3 + 1/2 M_{\rm ce}^2 N^2 + 1/4 M_{\rm ce} N$ CM.
The set of $N$ ECMs, defined in \eqref{eq_ECT_BEM_def}, (after applying the size reduction $M' \to M_{\rm b}$ according to \eqref{eq_DE_size_reduction} or \eqref{eq_DE_size_reduction_odd}) requires total $MN(M_{\rm h}Q + M_{\rm b}Q + M_{\rm b}N)$ CM. 
For the data estimator matrix \eqref{eq_UW_Omega_def}, we first require $C_{\rm Herm} = M_{\rm b} M(M+1)/2$ CM to calculate $\mathbf{\breve{H}}_n^{\rm H}\mathbf{\breve{H}}_n$, followed by the Cholesky decomposition of $(\mathbf{\breve{H}}_n^{\rm H}\mathbf{\breve{H}}_n + \sigma_{w}^2\mathbf{I}_M )^{-1} \mathbf{\breve{H}}_n^{\rm H}$, which takes $C_{\rm inv} = M^3/6 + M^2M_{\rm b} + M M_{\rm b}$ CM \cite{b_Huemer_UWODFM_1}.
The data estimation path is described by \eqref{eq_UW_X_est_vec_def} and \eqref{eq_UW_x_est_DD_mat_def}, which require $M_{\rm b} M N + C_{\rm sft}$ CM.

\subsection{UW-OTFS}

On the Tx side of UW-OTFS, ISFT is performed, followed by precoding with $\mathbf{G}$, which requires $M_{\rm s} M N$ CM, and with IFFT of size $M'$, which requires $NC_{\rm ifft}$ CM per OTFS frame.
The channel estimation signal path on the Rx side begins by computing $\mathbf{\Omega}_{\rm ce}$ in \eqref{eq_UW_Omega_CE_def}, which follows the same formula as in CP-OTFS, only here $M_{\rm ce} = M_{\rm h}$.
The GCE-BEM coefficient matrix is then estimated using \eqref{eq_CE_LMMSE}, which requires $M_{\rm h}^2N^2$ CM, and then used to calculate the set of $N$ ECMs according to \eqref{eq_UW_est_ECM_def}, which requires total $M_{\rm b}N(M_{\rm h} + 2MN)$ CM. The data estimator matrix \eqref{eq_UW_Omega_def} and the rest of the data estimation are identical to that of CP-OTFS.
Table.~\ref{table:1} contains a summary of the CM expressions (note that in UW-OTFS, $M_{\rm ce} = M_{\rm h}$). 


\renewcommand{\arraystretch}{1.4}
\begin{table}[t]
\caption{Memory footprint Expressions}
\label{table_mem}
\centering
\begin{tabular} {|c|c|c|} 
\hline\hline
 {} &  {CP-OTFS} &  {UW-OTFS}\\ \hline\hline
Tx side & $MN + MQ$ & $MN + M_{\rm s}M + M'N$ \\ \hline
$\mathbf{A}_{\rm ce}$, $\mathbf{\Omega}_{\rm ce}$ & \multicolumn{2}{c|}{$M_{\rm h}M_{\rm ce}N^2$, $M_{\rm h}M_{\rm ce}N^2$}  \\ \hline
Tensors & $M_{\rm b}MNQ + M_{\rm h}MQ + N^2$ & $M_{\rm b}MN + M_{\rm b}M_{\rm h}N + N^2$ \\ \hline
$\mathbf{H}_{n}$, $\mathbf{\Omega}_{n}$ & \multicolumn{2}{c|}{$M_{\rm b}M$} \\ \hline
    \hline
\end{tabular}
\end{table}

\subsection{Memory Footprint Analysis}

When evaluating CM counts in the analysis above, we summarized the equivalent number of CM required for each signal processing task. This is different from the actual resource allocation in FPGA because in some cases a single complex multiplier instance handles many CM operations per given amount of time.  
For example, an FFT core of size 512 can calculate the result in:
\begin{itemize}
    \item 5120 cycles using 1 complex multiplier (radix-2)
    \item 1280 cycles using 3 complex multipliers (radix-4).
\end{itemize}
Typically, a non-pipelined FFT core of size $M$ and word length $k_{\rm w}$ bits requires a memory block of size $2k_{\rm w}M$ bits to contain the input/output samples and another block of size $k_{\rm w}M$ bits to store the twiddle factors, i.e., $3k_{\rm w}M$ bits of memory resources per FFT core. 
However, in practical applications, several instances of FFT are applied in parallel to find balance between the throughput/latency of the algorithms and the memory requirements. 
Since such an analysis is above the scope of this paper, we provide a simpler analysis, taking into account the vectors/matrices/tensors that require memory allocation and we leave out the signal processing algorithms requiring random access memory such as matrix multiplication, FFT, or Cholesky decomposition.
Table~\ref{table_mem} provides comparison of memory resources (in complex samples) between CP-OTFS and UW-OTFS. 
The ``Tx side" row consists of input m-QAM symbol matrix $\mathbf{D}$ and the pulse shaping vector $\boldsymbol{\psi}$ in the CP-OTFS case, and of $\mathbf{D}$ and the precoding matrix $\mathbf{G}$ in the UW-OTFS case.
The ``Tensors" row refers to total memory footprint of the offline-calculated tensors defined in \eqref{eq_Complexity_Gk_mat_def}, \eqref{eq_Complexity_Mk_mat_def} and \eqref{eq_UW_Complexity_N_mat_def} for CP-OTFS, and tensors defined in \eqref{eq_UW_Complexity_G_tensor_def}, \eqref{eq_UW_Complexity_M_tensor_def}, and \eqref{eq_UW_Complexity_N_mat_def} for UW-OTFS.

\section{Numerical Analysis}
In this section, we numerically analyze the upper bound of spectral efficiency and evaluate the computational complexity of both UW-OTFS and CP-OTFS for an example of system parameter settings.

\subsection{System Numerology}

Due to the fundamental differences in structures of UW-OTFS and CP-OTFS, it is very hard (if not impossible) to adjust their system parameters so that they match in the bandwidth and the delay-Doppler grid size.
In our best effort, we selected the following parameters of UW-OTFS (system A) and CP-OTFS (system B).

Both systems used $M' = 128$, $M=32$, $N=16$ and $\Delta f = 26$ kHz. 
The maximum delay spread of the LTV channel was $\tau_m = 2.5 \ \mu s$, so the corresponding number of oversampled delay taps, identical for both systems, was $L' = \nint{\tau_m M' \Delta f} +1 = 9$. 
The GI length in UW-OTFS was set to $M_{\rm gi} = 2L' - 1 = 17$ to satisfy \eqref{eq_UW_Channel_Memory}, and the CP length in CP-OTFS was set to $M_{\rm cp} = L' -1 = 8$ to satisfy \eqref{eq_CP_Channel_Memory}.

The maximum velocity was set to $\mathscr{v}_{\rm max} = \{ 200, 400\}\ \rm km/h$, which corresponds to the maximum Doppler spread $\nu_{\rm max} =  \{ 1853.1, \ 3406.2\} \ \rm Hz$ for the carrier frequency $f_0 = 10 \ \rm GHz$.
The corresponding Doppler spread in terms of Doppler grid taps for $\mathscr{v}_{\rm max} = \{ 200, 400\}\ \rm km/h$ was $\pm \{1.21, 2.23\}$ in CP-OTFS and $\pm \{1.14, 2.10\}$ in UW-OTFS.

The number of active subcarriers of UW-OTFS is determined by $M_{\rm s}^{(A)} = M+M_{\rm gi} = 49$, and in CP-OTFS we have $M_{\rm s}^{(B)} = M + 2M_\alpha$, where $M_\alpha = \floor{\alpha M/2}$.
The number of FD bins used for data estimation, $M_{\rm b}$, was set equal to $M_{\rm s}$ in both systems.

\renewcommand{\arraystretch}{1.15}
\begin{table}[t]
\caption{Numerology of UW-OTFS and CP-OTFS}
\label{table:2}
\centering
\begin{tabular} {|c|c|c|c|c|} 
\hline\hline
 {Parameter} &  {UW} &  {CP*} & {CP**} & {CP***} \\ \hline\hline
$\Delta f$ [kHz] & $26.0$  & $26.0$ & $26.0$ & $26.0$\\ \hline
$M$ [-]          & $32$    & $32$   & $32$   & $32$  \\ \hline
$M_g$, $M_{\rm ce}$, $m_{\rm ce}$ [-] & - & $9,6,2$ & $11,6,3$ & $9,6,2$\\ \hline
$M_{\alpha}$ [-] & -  & $6$ & $5$ & $8$\\ \hline
$M_{\rm s}$ [-] & $49$  & $44$ & $42$ & $48$\\ \hline
$BW$ [MHz] & $1.274$  & $1.144$ & $1.092$ & $1.248$\\ \hline
$T_{ \rm tx}$ [$\mu s$] & $615.4$  & $653.9$ & $653.9$ & $653.9$\\ \hline
$R_{\rm s}$ [kBd] & $832.0$ & $562.8$ & $513.8$ & $562.8$\\ \hline
$\eta$ [bit/s/Hz] & $2.612$  & $1.968$ & $1.882$ & $1.804$\\ 
    \hline\hline
\end{tabular}
\end{table}

\begin{figure}[t] \centering
\includegraphics[scale=0.40]{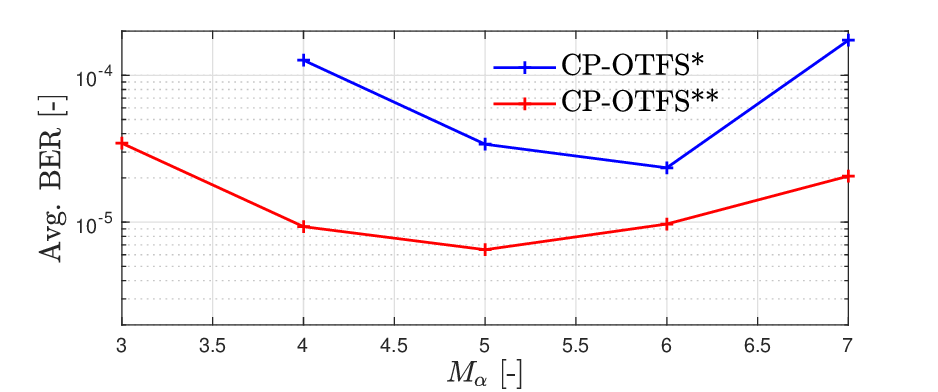}
\caption{Optimal $M_\alpha$ in CP-OTFS* and CP-OTFS** for $400$ km/h and $E_{\rm b}/N_0 = 28$ dB.}
\label{fig_M_alpha}
\end{figure}

\subsection{Spectral Efficiency}

The upper bound of spectral efficiency (i.e., in case of no errors) is given by $\eta = k_b R_{\rm s}/BW$, where the upper limit of Baudrate is given by $R_{\rm s}^{(A)} = M^{(A)}N/T_{\rm tx}^{(\rm A)}$ and $R_{\rm s}^{(B)} = (M^{(B)}-M_g)N/T_{\rm tx}^{(B)}$. 
OTFS transmission durations are defined by $T_{\rm tx}^{(A)} = N/\Delta f$ and $T_{\rm tx}^{(B)} = (M'+M_{\rm cp})N/M'/\Delta f$ and the approximate system bandwidth can be defined as $BW = \Delta f M_{\rm s}$. 
Lastly, the CP-OTFS pilot guard parameters of the delay domain, $M_{\rm g}$, $M_{\rm ce}$, and $m_{\rm ce}$, see Fig.~\ref{fig_DD_leakage}, had to be found.
After optimizating the spectral efficiency and BER, we found two suitable combinations of the parameter triplet $M_{\rm g},\ M_{\rm ce},\ m_{\rm ce}$ : 
\begin{itemize}
\item $9, 6 ,2$ with $M_\alpha = 6$, yielding $M_{\rm s} = 44$ (denoted as CP-OTFS*)
\item $11, 6, 3$ with $M_\alpha = 5$, yielding $M_{\rm s} = 42$ (denoted as CP-OTFS**).
\end{itemize}
Unfortunately, both combinations result in a slightly lower $M_{\rm s}$ than that of UW-OTFS; therefore, also a slightly smaller $BW = \Delta f M_{\rm s}$, ($10$ and $14$ percent for CP-OTFS* and CP-OTFS**, respectively).
It is possible to further increase $M_\alpha$ of CP-OTFS and approach $BW$ of UW-OTFS, hence, we introduce CP-OTFS***, which is CP-OTFS* where $M_\alpha = 8$. However, as it is shown in Fig. \ref{fig_M_alpha} and also in the Simulations section, the increase of $M_\alpha$ leads to suboptimal BER performance.

The comparison of numerologies is summarized in Table~\ref{table:2}.
We can see that even with slightly higher $BW$, the achievable spectral efficiency of UW-OTFS, $\eta$, is approximately $36$ percent higher than that of CP-OTFS, i.e., $32.7$ percent higher than CP-OTFS* and $38.8$ percent higher than CP-OTFS**. 
This is due to the fact that UW-OTFS uses redundancy more efficiently than CP-OTFS.

It is worth noting that in CP-OTFS, we did not use any OOB reduction technique because it would further decrease its spectral efficiency.
For example, if we used the weighted overlap and add technique (WOLA) \cite{b_WOLA} - a popular choice in OFDM, another drop of $\eta$ would be inevitable, due to the increase in the size of the CP.

\renewcommand{\arraystretch}{1.15}
\begin{table}[t]
\caption{Computational Complexity Comparison}
\label{table:3}
\centering
\begin{tabular} {|c|c|c|} 
\hline\hline
 {} &  {CP-OTFS**} &  {UW-OTFS}\\ \hline\hline
Tx side & $10496$ & $34560$ \\ \hline
$\mathbf{\Omega}_{\rm ce}$ & $1331736$  & $2996388$ \\ \hline
$\mathbf{\tilde{H}}_{\rm ce}$ & $20736$  & $20736$ \\ \hline
$\mathbf{\tilde{H}}_{n}$ (all $N$) & $448512$    & $1057792$ \\ \hline
$\mathbf{\Omega}_{n}$ (all $N$) & $1151824$  & $1709392$ \\ \hline
Rx side & $2987112$  & $5826548$ \\ \hline
    \hline
\end{tabular}
\end{table}

\renewcommand{\arraystretch}{1.4}
\begin{table}[t]
\caption{Memory footprint Comparison}
\label{table_mem_comp}
\centering
\begin{tabular} {|c|c|c|} 
\hline\hline
 {} &  {CP-OTFS**} &  {UW-OTFS}\\ \hline\hline
Tx side & $640$ & $4128$ \\ \hline
$\mathbf{A}_{\rm ce}$, $\mathbf{\Omega}_{\rm ce}$ & $13824$ & $20736$ \\ \hline
Tensors & $87424$ & $42240$ \\ \hline
$\mathbf{H}_{n}$, $\mathbf{\Omega}_{n}$ & $1344$ & $2048$ \\ \hline
    \hline
\end{tabular}
\end{table}

\subsection{Evaluating the Computational Complexity and Memory Footprint}

Here, we evaluate the CM count expressions and memory footprint derived in Section VIII for the aforementioned numerology of CP-OTFS** and UW-OTFS to provide an example of a numerical comparison of both systems.

The numerical values of the CM counts are provided in Table~\ref{table:3}. These results indicate that in both systems, the Rx CM count is much higher than the Tx CM count (approximately 168 times higher for UW-OTFS and 285 times higher for CP-OTFS). 
The main contributors to Rx complexity are related to the computation of $\mathbf{\Omega}_{\rm ce}$, $\mathbf{\widetilde{H}}_{n}$, and $\mathbf{\Omega}_{n}$.  
The total CM count for the Rx side of UW-OTFS is $1.95$ times higher than that of CP-OTFS, which is mainly because $M_{\rm h} > M_{\rm ce}$, ($M_{\rm h} = 9$ and $M_{\rm ce} = 6$), and also because $M_{\rm s}$ is smaller in CP-OTFS due to the selected system numerology.

Table~\ref{table_mem_comp} shows the numerical comparison of the memory footprint in complex samples between CP-OTFS** and UW-OTFS. The offline-calculated tensors consume most of the memory resources, however, further optimizations can be applied and some of the memory content can be calculated in the real time, reducing the memory footprint for the price of increased computational complexity.

\section{Simulations}

This section contains the Monte Carlo simulation results of CP-OTFS and UW-OTFS. 
Both systems are compared in terms of BER and normalized mean squared error (NMSE) for a given $E_{\rm b}/N_0$ for several different settings of the LTV channel.
We used the numerology defined in Section IX and the input symbols were drawn from a 16-QAM constellation without applying any forward error code.
We used the LTV channel model in \eqref{eq_yk_channel_def} with $P = 16$ sparse taps and the random channel parameters were generated as follows:
\begin{itemize}
\item $k_{{\tau'}}[i]$ was uniformly distributed from within $[ 0, L'-1 ]$
\item $\xi[i]$ used the Jakes' distribution with range $[ -\nu_{\rm max}, \nu_{\rm max})/\Delta f$
\item $h[i]$ was complex-normal distributed with zero mean and unit variance. Uniform power delay profile was used.
\end{itemize}

Monte Carlo simulations of BER and NMSE for each $E_{\rm b}/N_0$ point were performed across $4000$ random channel realizations. Each simulation run consisted of the following steps:
\begin{enumerate}
    \item parameters $k_{{\tau'}}[i]$, $\xi[i]$, and $h[i]$ are generated randomly according to their statistics. 
    \item the oversampled CP-OTFS signal, as defined in \eqref{eq_mCP_TX_s_k_PS_def}, and UW-OTFS signal from \eqref{eq_TX_s_0_waveform_1} are generated and propagated through the LTV channel model with input/output relation \eqref{eq_yk_channel_def}.
    For the UW-OTFS signal, three options are considered: the proposed UW pilot \eqref{eq_TX_s_u_waveform_1}, the Dirac impulse pilot \eqref{eq_TX_s_u_ideal_waveform_1} denoted by $\delta$, and no pilot in the case of perfect CSI.
    For CP-OTFS, no pilot is used in the perfect CSI case.
    \item average Rx input signal power $P_{\rm s}$ is measured (i.e., the LTV channel output before AWGN addition) and the AWGN variance, used for AWGN generation in the simulation, is calculated as
\begin{equation}\label{eq_UW_AWGN_EbN0_def}
     \sigma_{\rm w}^2 = \frac{1}{k_{\rm bps} } \frac{P_{\rm s}}{E_{\rm b}/N_0[-]}.
    \end{equation}
     Here, $k_{\rm bps}^{(A)} = k_{\rm b} M/M'$ is the number of bits per oversampled delay domain sample for UW-OTFS and $k_{\rm bps}^{(B)} = k_{\rm b} (MN-M_0)/N/M'$ is the same for CP-OTFS, where $k_{\rm b} = 4$ for the 16-QAM constellation.
     \item AWGN is added to the received signal. 
     \item Rx processing algorithms (i.e., CSI estimation and data estimation) produce the data symbol estimates followed by the calculation of BER and NMSE.
\end{enumerate}
The NMSE is calculated from the perfect CSI and estimated CSI ECMs according to
\begin{equation}\label{eq_NMSE_def1}
J = \frac{1}{N}  \sum_{n=0}^{N-1}  \frac{\big(\mathbf{{\tilde{h}}}_n -  \mathbf{h}_n\big)^{\rm H}\big( \mathbf{{\tilde{h}}}_n -  \mathbf{h}_n\big)}{\mathbf{h}_n^{\rm H}\mathbf{h}_n}.
\end{equation}
Here, $\mathbf{h}_n$ and $\mathbf{\tilde{h}}_n$ denote the reshaped ECMs for the perfect and estimated CSI, respectively, defined as $\mathbf{h}_n \equiv \vect\{\mathbf{H}_n\}$ and $\mathbf{\tilde{h}}_n \equiv \vect\{\mathbf{\widetilde{H}}_n\}$.
Throughout the simulation, the CP-OTFS's pilot energy, $\rho_0^2$, matches the equivalent of the UW energy in UW-OTFS, $\sigma_{\rm u}^2$, so that the following formula holds true
\begin{equation}\label{eq_UW_CP_pilot_energy_def}
 \rho_0^2 = \frac{\sigma_{\rm u}^2}{1-\sigma_{\rm u}^2 } \Big(\frac{MN}{M_0} - 1\Big).
 \end{equation}

\subsection{Average BER and NMSE Comparison}

\begin{figure}[t] \centering
\begin{tabular}{c} 
\includegraphics[scale=0.45]{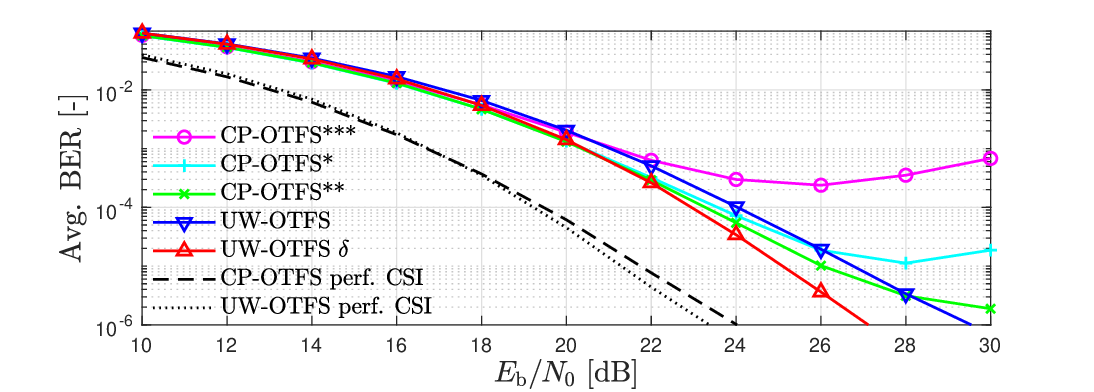}\\ \includegraphics[scale=0.45]{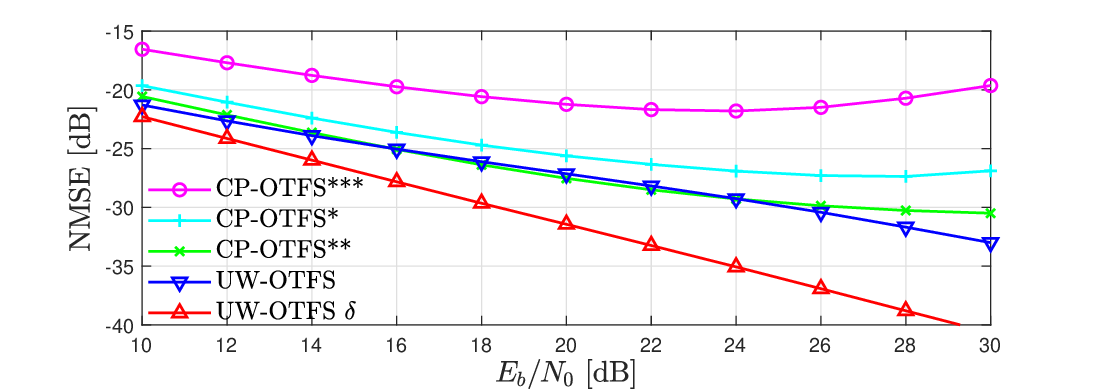}
\end{tabular} 
\caption{Average BER (top) and NMSE (bottom) of UW-OTFS and CP-OTFS for $200 \ \rm km/h$.}
\label{fig_ABER_200kmph}
\end{figure}

\begin{figure}[t] \centering
\begin{tabular}{c} 
\includegraphics[scale=0.45]{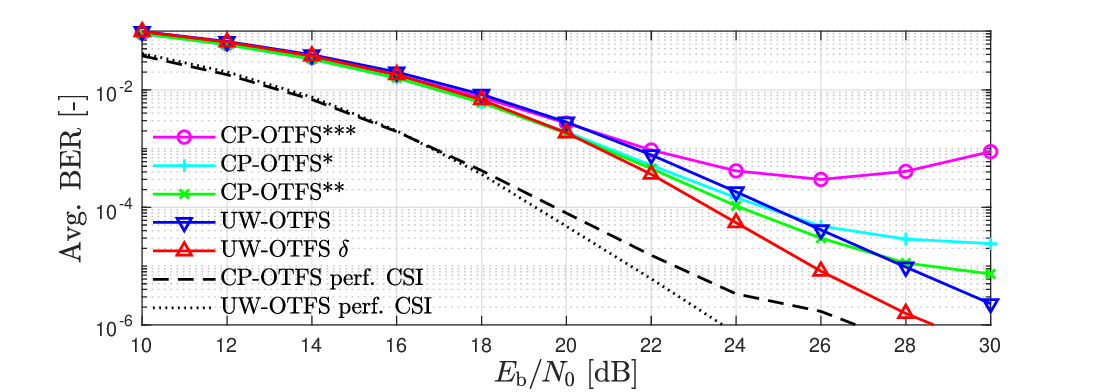}\\ \includegraphics[scale=0.45]{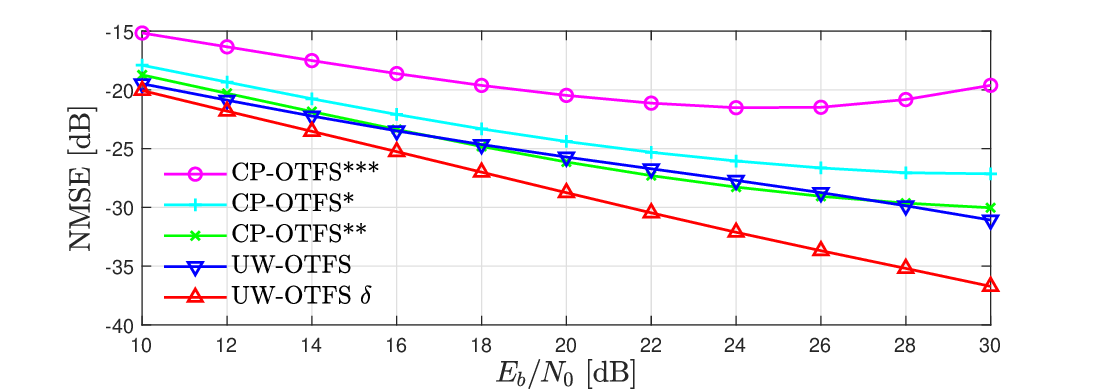}
\end{tabular}
\caption{Average BER (top) and NMSE (bottom) of UW-OTFS and CP-OTFS for $400 \ \rm km/h$.}
\label{fig_ABER_400kmph}
\end{figure}

In the upper parts of Fig.~\ref{fig_ABER_200kmph} and Fig.~\ref{fig_ABER_400kmph} we see an average BER comparison between UW-OTFS and the two CP-OTFS numerologies for $200\ \rm km/h$ and $400\ \rm km/h$, respectively.
First, we may notice the difference between the Dirac UW (marked by $\delta$) and the proposed UW pilot. This performance loss represents the price for better PSD properties of the proposed UW pilot, as discussed in Section VI-C.

Next, we compare CP-OTFS* and CP-OTFS**. 
We see that CP-OTFS* experiences a higher error floor than CP-OTFS** in both $200\ \rm km/h$ and $400\ \rm km/h$. 
This is because CP-OTFS* uses a smaller $M_{\rm g}$, which causes higher data-to-pilot energy leakage, thus damaging the CSI estimate.
After observing the NMSE in the bottom parts of Fig.~\ref{fig_ABER_200kmph} and Fig.~\ref{fig_ABER_400kmph}, we see the saturation in the high $E_{\rm b}/N_0$ regions of CP-OTFS*, which confirms that the problem lies in the corrupted CSI estimate.
If we now compare UW-OTFS with the proposed UW pilot to CP-OTFS, we can see that in the high $E_{\rm b}/N_0$ region, UW-OTFS outperforms both CP-OTFS* and CP-OTFS**. This is due to the absence of the energy leakage issue in the UW-OTFS.

CP-OTFS*** shows a notable decrease in ABER and NMSE performance, which is due to the use of higher $M_\alpha$ than the optimum depicted in Fig.~\ref{fig_M_alpha}. 
This performance loss occurs despite having a higher roll-off factor, therefore, it is not linked to the delay energy leakage phenomenon. It is explained in detail in Section III-B of \cite{bayat2023unified}.

Lastly, in the perfect CSI case, we also observe a performance advantage of UW-OTFS, which could be attributed to the precoding matrix $\mathbf{G}$, which acts like a Reed-Solomon code in the complex domain \cite{b_UWOFDM_RScode}.

 \subsection{Optimizing the Pilot Energy}

\begin{figure}[t] \centering
\includegraphics[scale=0.45]{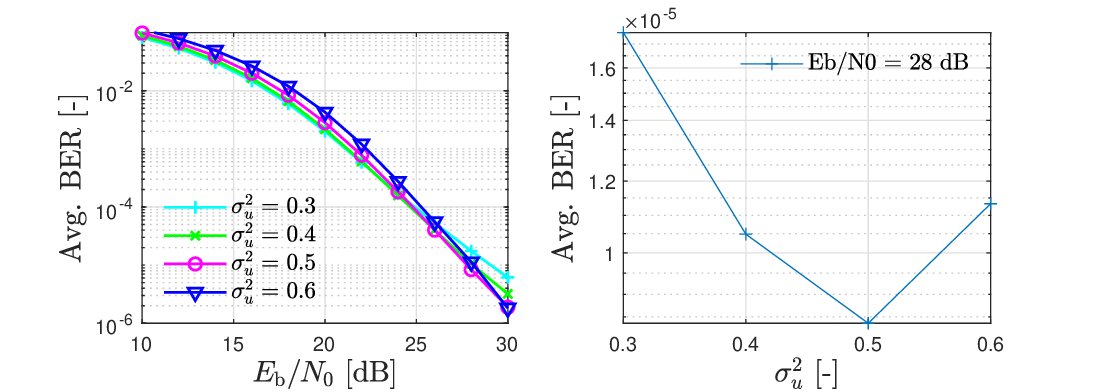}
\caption{Average BER of UW-OTFS at $400 \ \rm km/h$ and for different values of $\sigma_u^2$ (left) and a detail for $E_{\rm b}/N_0 = 28$ dB (right).}
\label{fig_Opt_PER}
\end{figure}

The pilot energy $\sigma_u^2$ strongly influences the performance of both UW-OTFS and CP-OTFS. If set too low, the CSI estimate is noisy, whereas if $\sigma_u^2$ is set too high, the data energy is reduced accordingly, which leads to suboptimal average BER performance. Therefore, $\sigma_u^2$ was set to $0.5$, as determined from the numerical analysis presented in Fig.~\ref{fig_Opt_PER}. The CP-OTFS pilot energy was then set according to \eqref{eq_UW_CP_pilot_energy_def}.

 \subsection{Pilot Cancellation Analysis}

\begin{figure}[t] \centering
\includegraphics[scale=0.50]{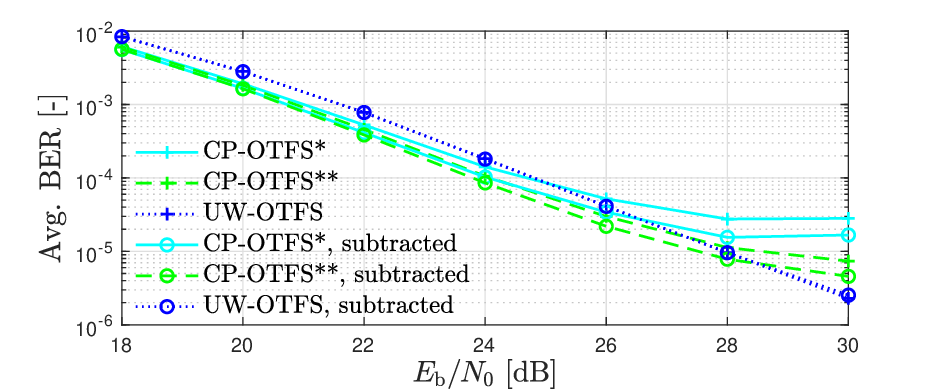}
\caption{The influence of pilot cancellation in CP-OTFS and UW-OTFS for $400 \ \rm  kmph$.}
\label{fig_ABER_NSUB_PSUB}
\end{figure}

Since both UW-OTFS and CP-OTFS use pilots that interfere with the data signal, a logical question would be whether we should subtract the pilot on the Rx side before the data estimation. 
Practically, this could be done by first estimating the CSI and then using it to distort the pilot, which could then be subtracted from the received signal. Alternatively, we could use various iterative methods for this task.
In Fig.~\ref{fig_ABER_NSUB_PSUB}, we analyze whether there is any benefit of subtracting a pilot that was distorted by a perfectly known CSI. 
It is clear that in UW-OTFS, the effect of pilot cancelation is rather marginal.
This is mainly due to the UW pilot's centering in the middle of GI, which is ignored by the data estimation algorithm. 
In CP-OTFS, we can see a slightly more significant benefit of pilot cancelation; however, the main cause of the error floor in CP-OTFS is the data-to-pilot energy leakage which cannot be treated by pilot cancelation.

\subsection{Optimizing the BEM Rate}

\begin{figure}[t] \centering
\includegraphics[scale=0.50]{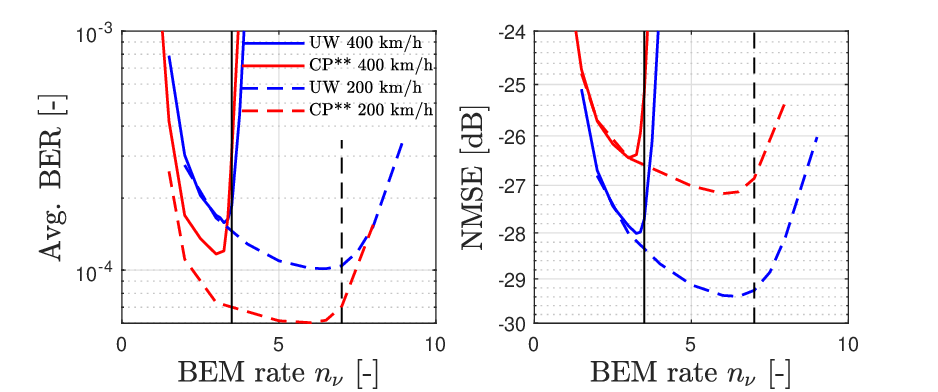}
\caption{Search for the optimal BEM rate in UW-OTFS and CP-OTFS** for $E_{\rm b}/N_0 = 24 $ dB and $200 \ \rm  km/h$ and $400 \ \rm  km/h$ in terms of average BER (left) and NMSE (right). The black vertical lines represent the values of \eqref{eq_optimal_BEM_rate}, i.e. $n_\nu = 3.5$ and $n_\nu = 7.0$, respectively. }
\label{fig_BEM_rate}
\end{figure}

As introduced in Section VI-B, the GCE-BEM rate, $n_\nu$, can be adjusted according to \eqref{eq_optimal_BEM_rate} to achieve better BER performance. 
In Fig.~\ref{fig_BEM_rate} we can observe the influence of the BEM rate on average BER and NMSE for $E_{\rm b}/N_0 = 24 \ dB$ and for two different velocities, $200 \ \rm km/h$ and $400 \ \rm km/h$. 
It is certain that precise adjustments of $n_\nu$ based on the expected maximum Doppler spread $\nu_{\rm max}$ according to \eqref{eq_optimal_BEM_rate} are beneficial to both UW-OTFS and CP-OTFS. 
The simulations in Fig.~\ref{fig_ABER_200kmph} and Fig.~\ref{fig_ABER_400kmph} use BEM rates $7.0$ and $3.5$, respectively for UW-OTFS and $6.0$ and $3.0$, respectively for CP-OTFS.

\begin{figure}[t] \centering
\begin{tabular}{c} 
\includegraphics[scale=0.45]{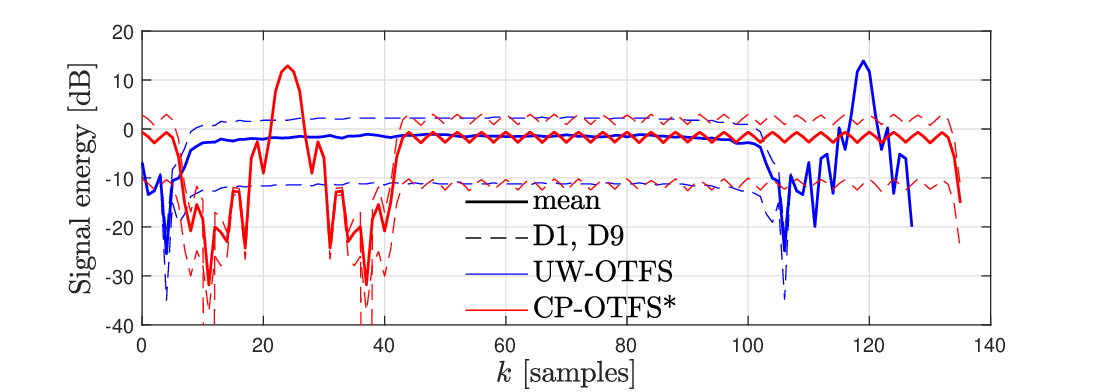}\\ 
\begin{tabular}{c c} \includegraphics[scale=0.45]{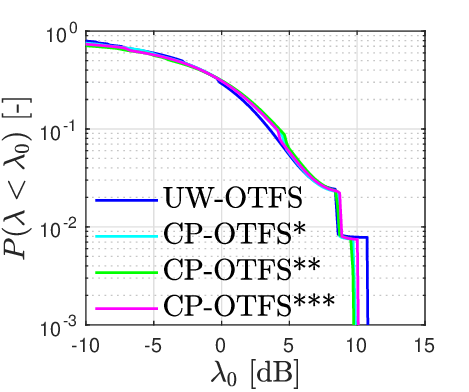} & \includegraphics[scale=0.45]{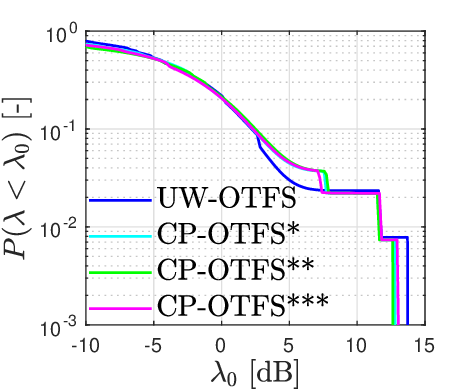}
\end{tabular}
\end{tabular}
\caption{Energy profile of a UW-OTFS and CP-OTFS* delay block (top) and PAPR comparison between all numerologies for $\sigma_u^2 = 0.25$ (bottom left) and $\sigma_u^2 = 0.5$ (bottom right).}
\label{fig_Energy_PAPR}
\end{figure}

\subsection{PAPR Analysis}

In the top part of Fig.~\ref{fig_Energy_PAPR}, we can see a statistical energy profile of a single UW-OTFS and CP-OTFS* delay block. 
It is clear that maximum energy levels are in both cases defined by the pilots, which are deterministic, i.e., their decile curves (D1, D9) unite with the mean energy curve, unlike the stochastic parts of the delay block.
The dominant pilot is the reason why the PAPR complementary cumulative distribution function (CCDF) in the bottom part of Fig.~\ref{fig_Energy_PAPR} contains sharp edges. 
After observing UW-OTFS and all three numerologies of CP-OTFS, we can see that PAPR of UW-OTFS is at most 1.1 dB higher than CP-OTFS*.

  \section{Conclusion}

In this paper, we elaborate the issue of delay domain energy leakage caused by oversampling and pulse shaping in OTFS with embedded pilot. 
We first analyze the issue theoretically using a design we denote as CP-OTFS, which suffers interference due to the leakage of data energy to the channel estimate. 
Then we present a novel system called UW-OTFS that is free of the interference, because its pilot is oversampled by design and does not undergo leakage.
Both systems are equipped with a novel GCE-BEM-based channel estimation technique that allows the utilization of LMMSE for both channel estimation and data estimation.
Monte Carlo simulations in the fractional Doppler LTV channel reveal that UW-OTFS is free of the error floor, which is present in CP-OTFS due to interference. 
Moreover, UW-OTFS shows a 36-percent increase in achievable spectral efficiency - compared to CP-OTFS - because it distributes redundancy more carefully.
Lastly, we analyze the computational complexity of both systems in terms of the complex multiplier equivalents and memory footprint.
The complexity of UW-OTFS is greater than that of CP-OTFS, mainly due to differences in the selected system numerology. 
Comparison of UW-OTFS with other channel estimation methods with oversampling and pulse shaping is an interesting future work.

\section*{Acknowledgments}
We thank Oliver Lang from Johannes Kepler University, Linz, Austria, and Christian Hofbauer from Silicon Austria Labs, Linz, Austria, for technical discussions on the concept of UW-OFDM.

\appendices
\numberwithin{equation}{section}

\section{Detailed Derivation of the Leakage Analysis}

The ECM in \eqref{eq_mCP_ECT_PS_def}, after assuming $P=1$, delay $k_{\tau'0}$ and zero Doppler frequency, is simplified to
\begin{equation}\notag
H_n[m, m'] =   \sqrt{Q}  h_0    e^{-j2\pi  \frac{ k_{\tau'0} m'}{M Q} } 
\end{equation}
\begin{equation}\label{eq_A1_ECM_1}
\times \sum_{k = 0}^{Q-1} e^{-j2\pi  \frac{ k_{\tau'0} k}{Q} } \psi[m'+kM]  \delta \big[ (m'+kM - m)_{M'} \big].
\end{equation}
The FT domain matrix in \eqref{eq_mCP_Y_mat_eval_4} is then expanded as
\begin{equation}\notag
Y[m,n] = \frac{x_0 h_0 \sqrt{Q}}{\sqrt{MN}} e^{j2\pi \frac{nq_0}{N}} \sum_{m'=0}^{M-1} e^{-j2\pi  \frac{ k_{\tau'0} m'}{M Q} } e^{-j2\pi \frac{m' p_0}{M}}  
\end{equation}
\begin{equation}\label{eq_A1_Y_1}
\times \sum_{k = 0}^{Q-1} e^{-j2\pi  \frac{ k_{\tau'0} k}{Q} } \psi[m'+kM]  \delta \big[ (m'+kM - m)_{M'} \big].
\end{equation}
Next, the Rx filtering and aliasing is performed by $\mathbf{\hat{Y}} = \mathbf{A}_{\rm d}^{\rm T} \boldsymbol{\Psi} \mathbf{Y}$.
Using the following identity,
\begin{equation}\notag
\sum_{m=0}^{QM-1} f[m] = \sum_{k'=0}^{Q-1} \sum_{m=0}^{M-1} f[m + k'M],
\end{equation}
we may evaluate the elements of $\mathbf{\hat{Y}}$ as follows,
\begin{equation}\notag
\hat{Y}[p,n] = \sum_{k'=0}^{Q-1}  \psi[p + k'M] Y[p + k'M,n],
\end{equation}
and after including \eqref{eq_A1_Y_1}, it is expanded into
\begin{equation}\notag
\hat{Y}[p,n] = \frac{x_0 h_0 \sqrt{Q}}{\sqrt{MN}} e^{j2\pi \frac{nq_0}{N}}     \sum_{m'=0}^{M-1}  e^{-j2\pi \frac{m' (p_0 + k_{\tau'0}/Q)}{M}}
\end{equation}
\begin{equation}\notag
\times \sum_{k = 0}^{Q-1} e^{-j2\pi  \frac{ k_{\tau'0} k}{Q} }  \psi[m'+kM]  
\end{equation}
\begin{equation}\label{eq_A1_Y_2}
\times \sum_{k'=0}^{Q-1}  \psi[p + k'M]  \delta \big[ (m'-p + M(k-k') )_{M'} \big].
\end{equation}
After considering the ranges of $m', p, k, k'$, we conclude that $\delta \big[ (m' -p + M(k- k'))_{M'} \big] = \delta \big[ m' -p + M(k- k') \big]$, so the only case that produces a nonzero result is when $m' = p$ and $k = k'$. 
Therefore, \eqref{eq_A1_Y_2} obtains a more compact form,
\begin{equation}\label{eq_A1_Y_3}
\hat{Y}[p,n] = \frac{x_0 h_0 \sqrt{Q}}{\sqrt{MN}} e^{j2\pi \frac{nq_0}{N}}    e^{-j2\pi \frac{p (p_0 + k_{\tau'0}/Q)}{M}} \hat{\psi}[p],  
\end{equation}
where $\hat{\psi}[p]$ is defined in \eqref{mCP_Xi_vec_def}.
After transforming \eqref{eq_A1_Y_3} into the DD domain with SFT, we obtain \eqref{eq_mCP_y_DD_mat_final}.
Now, when $k_{\tau'0} = 0$, we may observe the following property of a Nyquist filter, 
\begin{equation}\label{eq_A1_y_DD_mat_eval}
 \hat{\psi}[l] = \sum_{k' = 0}^{Q-1}  \psi^2[l + k'M] = 1, \quad \forall \quad l \in \{0,1,\dots, M-1\},
\end{equation}
which further simplifies \eqref{eq_mCP_y_DD_mat_final} to \eqref{eq_mCP_y_DD_mat_final_even} for the integer values of downsampled delay taps.


\end{document}